\documentclass[journal=jacsat,layout=twocolumn]{achemso}

\usepackage{chemformula} 
\usepackage[T1]{fontenc} 
\usepackage{graphicx}
\usepackage{dcolumn}
\usepackage{bm}
\usepackage{multirow} 
\usepackage{booktabs}
\usepackage{times,mathptmx}
\usepackage{makecell}
\usepackage[colorlinks=true,citecolor=blue,linkcolor=blue]{hyperref}

\makeatletter
\def\blfootnote{\xdef\@thefnmark{}\@footnotetext}
\makeatother

\makeatletter

\author{Johnathan Kowalski}
\affiliation[ORNL]{Center for Nanophase Materials Sciences, Oak Ridge National Laboratory, Oak Ridge, Tennessee 37831, United States}
\alsoaffiliation{ORISE Research Program Participant in the SULI Program sponsored by DOE WDTS and hosted at ORNL}

\author{Liangbo Liang}
\affiliation[ORNL]{Center for Nanophase Materials Sciences, Oak Ridge National Laboratory, Oak Ridge, Tennessee 37831, United States}
\email{liangl1@ornl.gov}

\title[Raman Digital Twin of Janus TMDs]{Raman Digital Twin of Monolayer Janus Transition Metal Dichalcogenides}


\begin{document}

\begin{abstract}
Monolayer transition metal dichalcogenides (TMDs) are a key class of two-dimensional (2D) materials with broad technological potential. Their Janus counterparts exhibit unique properties due to broken out-of-plane symmetry and further enrich the functionalities of TMDs. However, experimental synthesis and identification of Janus TMDs remain challenging. It is thus highly desirable to have a rapid, simple, and \textit{in situ} characterization technique to monitor, in real time, the conversion process from the parent to Janus structure. Raman spectroscopy stands out for such a task as it is a powerful, non-destructive, and very commonly used tool to characterize 2D materials both \textit{in situ} and \textit{ex situ}. To realize the full potential of Raman spectroscopy on rapid characterization of Janus TMDs, we present a computational “Raman digital twin” library for various monolayer Janus TMDs in both 2H and Td phases. We focus on group-6 TMDs: MoS$_2$, WS$_2$, MoSe$_2$, WSe$_2$, MoTe$_2$, WTe$_2$ and their Janus variants: MoSSe, MoSTe, MoSeTe, WSSe, WSTe, and WSeTe. Using first-principles density functional theory (DFT), we calculate their vibrational properties and predict distinct Raman fingerprints. These phonon and Raman signatures reflect each material's structural symmetry and atomic composition, enabling clear identification via Raman spectroscopy. Our theoretical work supports experimental efforts by providing benchmarks for material identification, structural analysis, and quality control. The computational library expedites the discovery and development of Janus 2D materials, facilitating tighter integration between theoretical predictions and experimental validation.
\end{abstract}

\section{Introduction}
The discovery of graphene began a scientific revolution, serving as the first stable, monolayer material with extraordinary properties \cite{novoselovElectricFieldEffect2004}. This kick-started the worldwide effort into two-dimensional (2D)
materials and their diverse fundamental physical phenomena and technologically relevant properties \cite{Butler2013,Bhimanapati2015}. One of the most prominent classes of 2D materials are the transition metal dichalcogenides (TMDs), which could become building blocks for the next-generation technologies ranging from flexible electronics and photodetectors to energy storage devices and electrocatalysts due to their tunable bandgaps and rich physics \cite{susarlaQuaternary2DTransition2017, makAtomicallyThinMoS2New2010, riis-jensenClassifyingElectronicOptical2019, manzeli2DTransitionMetal2017}.
As the search for more complex and exotic 2D structures continues, there is a pressing need for reliable methods of identifying and characterizing these materials.

Group-6 TMDs such as MoS$_2$, WS$_2$, MoSe$_2$, WSe$_2$, MoTe$_2$, and WTe$_2$ are the most extensively studied type of TMDs. They follow the $X - M - X$ sandwich structure, where $M$ is the transition metal atom and $X$ are the chalcogen atoms. While conventional TMDs have an out-of-plane mirror symmetry, an interesting subclass known as Janus TMDs emerges with the breaking of this symmetry. Janus TMDs are achieved when one chalcogen layer is replaced with a different chalcogen atom, resulting in an asymmetric $X - M - Y$ structure \cite{luJanusMonolayersTransition2017}.
This asymmetry proves to be very important, as it induces many novel physical properties not present in the parent structures. One result is a built-in dipole moment, which can lead to strong piezoelectricity, along with the Rashba effect arising from the broken inversion symmetry and large spin-orbit coupling \cite{riis-jensenClassifyingElectronicOptical2019, tang2DJanusTransition2022, dongLargeInPlaneVertical2017}. These properties open the door to applications in flexible nanogenerators, photocatalysis, valleytronic and spintronic devices, and asymmetric gas sensing, where out-of-plane polarity and charge separation play a key role \cite{samadiGroup6Transition2018a}. Due to the appealing functionalities, the synthesis of Janus TMDs has been a major focus in the field \cite{dengSynthesisJanusMoSSe2025, linLowEnergyImplantation2020}.
However, despite multiple promising synthesis methods such as hydrogen plasmas sputtering and pulsed laser deposition \cite{luJanusMonolayersTransition2017,linLowEnergyImplantation2020,harrisRealTimeDiagnostics2D}, the process remains very difficult and sensitive to experimental setup. It is plagued by the possibilities of incomplete conversion and creating 
alloyed domains as opposed to pristine Janus layers. To facilitate synthesis of Janus TMDs and determine whether a Janus monolayer is successfully formed, it is highly desirable to have a rapid, simple, and \textit{in situ} characterization technique to monitor, in real time, the conversion process from the parent structure to the Janus. 

Raman spectroscopy has emerged as an ideal tool for such a task. As a rapid, non-destructive optical approach, it offers a unique vibrational "fingerprint" that is highly sensitive to the crystal structure, symmetry, and chemical bonding of materials. In addition to identifying layer number, stacking pattern, and 
crystal phase \cite{zhang2015phonon,liang2017low}, Raman spectroscopy has been widely applied to probe strain, doping, defect density in 
monolayer TMDs, as well as to map spatial inhomogeneities using Raman imaging techniques \cite{parkin2016raman,mahjouri2016tailoring, congApplicationRamanSpectroscopy2020, 
wuSpectroscopicInvestigationDefects2017, velickyStrainChargeDoping2020, dadgarStrainEngineeringRaman2018}. In terms of the conversion from the parent to Janus TMDs, the change of the atomic composition and the inherent breaking of the out-of-plane mirror symmetry unavoidably alter the lattice dynamics and Raman spectra \cite{tang2DJanusTransition2022}. The precise positions and characteristics of these new or modified Raman peaks could therefore serve as signatures for confirming the successful synthesis of a Janus structure \cite{petricRamanSpectrumJanus2021}. However, the interpretation of Raman spectra can be complex and challenging, especially for novel materials like Janus TMDs that very often have not been synthesized and characterized previously. Without a known reference spectrum, it is nearly impossible for experimentalists to definitively assign observed peaks to a Janus structure versus other possibilities like defects, strain, alloyed domains or residual parent material.

To help mitigate this issue, we carry out first-principles density functional theory (DFT) calculations to generate a computational library, a "Raman digital twin", of vibrational and Raman fingerprints for key TMDs and their Janus counterparts. We focus on the group-6 Janus TMDs in both 2H and Td phases, including monolayers MoSSe, MoSTe, MoSeTe, WSSe, WSTe, and WSeTe, derived from their well-known parent compounds (MoS$_2$, WS$_2$, etc.), which have been explored for applications in optoelectronics, catalysis, and energy storage \cite{samadiGroup6Transition2018a}. The computational database of vibrational patterns and Raman spectra of the parent and Janus structures reveals distinct fingerprints for each material in either 2H or Td phase. It not only enables the identification of each parent and Janus structure, but also provides an atomic-scale picture about how the Raman modes evolve during the parent-to-Janus conversion process, something experiments cannot reveal. Therefore, this work offers a crucial theoretical resource that can directly support and accelerate experimental efforts in the synthesis, identification, and characterization of these novel 2D materials.

\section{Computational Methods}

First-principles DFT calculations were performed using the 
Vienna \textit{Ab initio} Simulation Package (VASP) \cite{kresseInitiomoleculardynamicsSimulationLiquidmetalamorphoussemiconductor1994}.
The interaction between core and valence electrons was described using the projector augmented-wave (PAW) method \cite{blochlProjectorAugmentedwaveMethod1994, kresseUltrasoftPseudopotentialsProjector1999}.
To provide a concrete analysis we employed two common exchange-correlation functionals: the local density approximation (LDA)
\cite{perdewSelfinteractionCorrectionDensityfunctional1981} and the generalized gradient approximation (GGA) with the Perdew-Burke-Ernzerhof (PBE) 
flavor \cite{perdewGeneralizedGradientApproximation1996}. Using both functionals in tandem allows for a more robust analysis, as LDA tends to underestimate lattice constants, resulting in overbinding and higher vibrational frequencies, while PBE often overestimates lattice constants, leading to softer force constants and lower vibrational frequencies. These functionals are routinely used for
analysis on parent and Janus TMDs \cite{harrisRealTimeDiagnostics2D, linLowEnergyImplantation2020, xuPhononSpectrumElectronic2021, zengElectronicStructureElastic2015, dongDefectsDopingdrivenModulation2022} 
and provide a range where the real structural parameters and vibrational frequencies are located.

The electronic wavefunctions were expanded in a plane-wave basis with a cutoff energy of 350 eV. All structures were placed in 
a periodic unit cell with a vacuum layer of at least 20 \AA{} in the z-direction to prevent spurious interactions with 
replicas. For structural relaxation, the lattice vectors and atomic positions were fully relaxed until the Hellmann-Feynman forces 
on each atom were less than 0.001 eV/\AA{} and the total energy converged to within 10$^{-8}$ eV. The Brillouin zone was sampled
using a Monkhorst-Pack 24 $\times$ 24 $\times$ 1 k-point mesh for the materials most stable in the hexagonal 2H phase and a 24 $\times$ 14 $\times$ 1 
k-point mesh for the orthorhombic Td phase.

Phonon calculations were done using the finite displacement method from the Phonopy package \cite{togoImplementationStrategiesPhonopy2023}.
Supercells of $4 \times 4 \times 1$ for 2H systems and $3 \times 2 \times 1$ for Td systems were created. Both positive and negative atomic displacements ($\Delta = 0.03 \text{\AA}$) were introduced to the supercells for static calculations by VASP to obtain the forces, which were processed by Phonopy to construct the dynamic matrix. The diagonalization of the dynamic matrix provides phonon frequencies and phonon eigenvectors. 

For the final Raman intensity calculations, we utilized SpectroPy, an open-source Python package developed for this work and available on \href{https://github.com/TheorySpectroPy/SpectroPy}{GitHub}. The Raman activity corresponding to each phonon mode $j$ was determined from the $3 \times 3$ 
Raman tensor, $\mathbf{R}_{j}$. The components of the tensor, $R_{j, \alpha \beta}$, represent the first-order derivative of the macroscopic electronic 
polarizability tensor, $\chi_{\alpha \beta}$, with respect to the normal mode coordinate of the phonon according to Placzek approximation that has been applied to various 2D materials \cite{kongFirstprinciplesStudyMagnetoRaman2024, liangFirstprinciplesRamanSpectra2014,Corro2016,Miranda2017}. For both positive and negative atomic displacements along each direction in the unit cell (displacement amplitude = 0.03 $\text{\AA}$), the frequency-dependent polarizability (or dielectric) tensor is computed by VASP (LOPTICS = .TRUE.) and then the derivatives of the polarizability tensor are obtained via the finite difference scheme. For all calculations, a common laser excitation energy of 1.96 eV (632.6 nm) was used (i.e., the polarizability tensor at 1.96 eV was adopted). The derivatives of the polarizability tensor, along with the phonon frequencies and phonon eigenvectors, generate the Raman tensor for each phonon mode \cite{kongFirstprinciplesStudyMagnetoRaman2024, liangFirstprinciplesRamanSpectra2014,Corro2016,Miranda2017}. The Raman intensity, $I_{j}$, for a given laser scattering geometry is proportional to the squared projection of the Raman tensor $\mathbf{R}$ onto the incident ($\mathbf{e}_{i}$) and scattered 
($\mathbf{e}_{s}$) light polarization vectors: $I_j \propto | \mathbf{e}_s \cdot \mathbf{R}_j \cdot \mathbf{e}_i |^2$.
From this, we allow the simulation of more experimental setups with different polarization configurations, such as parallel (XX)
and cross (XY) where $I \propto | R_{11}|^2 $ and $I \propto | R_{12}|^2 $ respectively.

\section{Results and Discussion}

\begin{figure*}[ht!]
    \centering
    \includegraphics[width=0.7\textwidth]{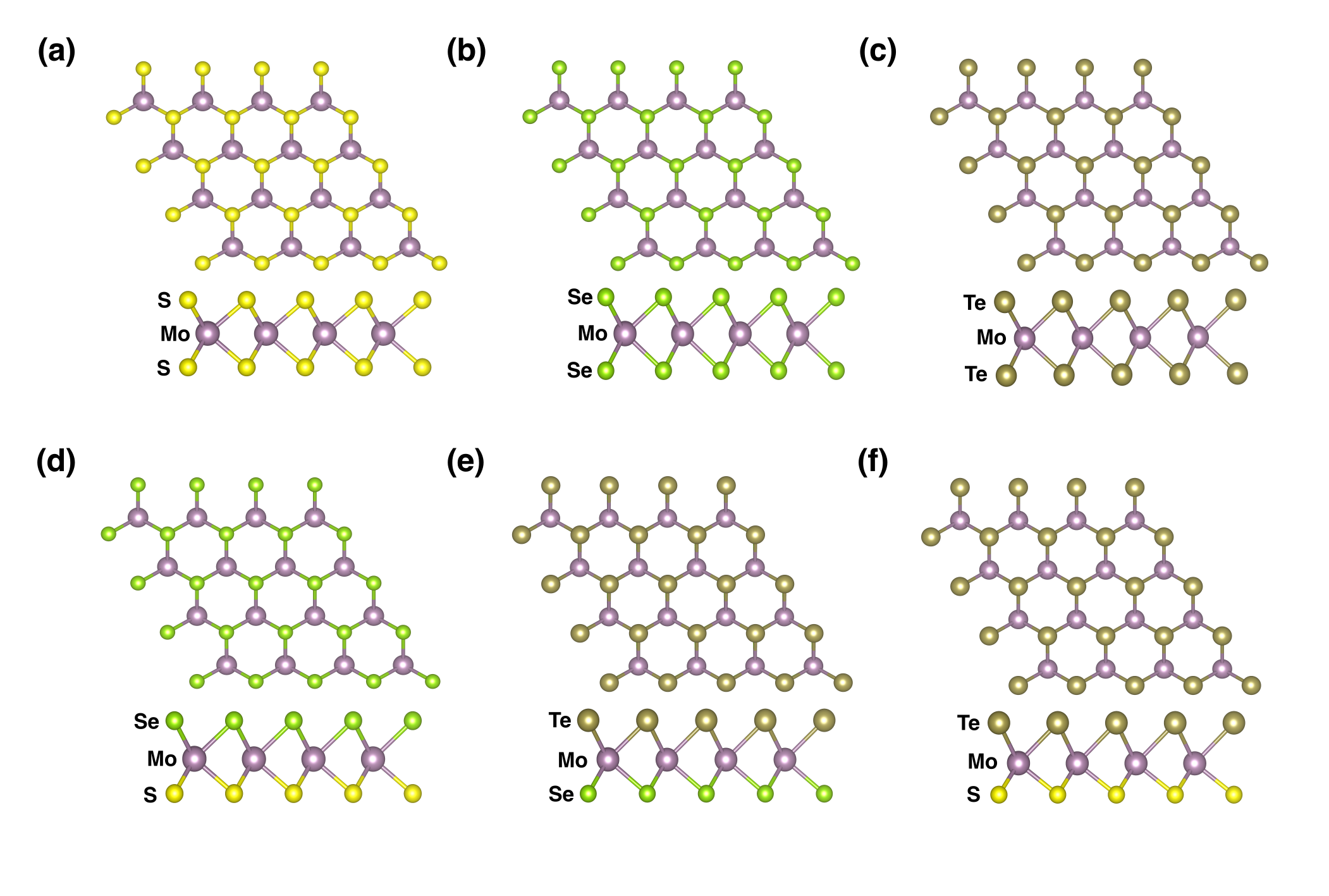}
    \caption{Side and top views of the optimized crystal structures of 
    (a) MoS$_2$, (b) MoSe$_2$, (c) MoTe$_2$, (d) MoSSe, (e) MoSeTe, and (f) MoSTe monolayers in the 2H phase.}
    \label{fig:mo-based-structure-labels}
\end{figure*}

\begin{table}[ht!]
    \centering
    \caption{Calculated lattice parameters for 2H-phase TMDs using LDA and PBE functionals.}
    \small
    \begin{tabular}{l|cc|cc}
        \hline
        \multirow{2}{*}{Material} & \multicolumn{2}{c|}{LDA} & \multicolumn{2}{c}{PBE} \\
        & $a$ (\AA) & $b$ (\AA) & $a$ (\AA) & $b$ (\AA) \\
        \hline
        MoS$_2$   & 3.1197 & 3.1197 & 3.1825 & 3.1825 \\
        MoSe$_2$  & 3.2459 & 3.2459 & 3.3180 & 3.3180 \\
        MoTe$_2$  & 3.4672 & 3.4672 & 3.5502 & 3.5502 \\
        MoSSe     & 3.1906 & 3.1906 & 3.2497 & 3.2497 \\
        MoSeTe    & 3.3571 & 3.3571 & 3.4342 & 3.4342 \\
        MoSTe     & 3.2930 & 3.2930 & 3.3647 & 3.3647 \\
        WS$_2$    & 3.1319 & 3.1319 & 3.1821 & 3.1821 \\
        WSe$_2$   & 3.2532 & 3.2532 & 3.3162 & 3.3162 \\
        WTe$_2$   & 3.4739 & 3.4739 & 3.5524 & 3.5524 \\
        WSSe      & 3.1917 & 3.1917 & 3.2574 & 3.2574 \\
        WTeSe     & 3.3620 & 3.3620 & 3.4328 & 3.4328 \\
        WSTe      & 3.3004 & 3.3004 & 3.3668 & 3.3668 \\
        \hline
    \end{tabular}
    \label{tab:lattice-params}
\end{table}

We begin our analysis by focusing on the parent and Janus TMDs in the hexagonal 2H phase. The majority of TMDs are energetically most stable in this phase.
The optimized equilibrium lattice parameters for these structures, calculated using both LDA and PBE functionals, are presented 
in Table \ref{tab:lattice-params}. A representative view of the Mo-based 2H atomic structures are shown in Figure \ref{fig:mo-based-structure-labels}, with the W-based 
2H structures following the same trend. Our calculated lattice parameters for monolayer MoS$_2$, a benchmark system, are $a = b \approx 3.12$~\AA{}
with the LDA functional and $a = b \approx 3.18$~\AA{} with the PBE functional, which is in good agreement with the established values 
of prior works \cite{molina-sanchezPhononsSinglelayerFewlayer2011, mahjouri-samaniPatternedArraysLateral2015,luanFirstprinciplesStudyElectronic2017, liangFirstprinciplesRamanSpectra2014}. This confirms the validity of our structural relaxation approach, and provides confidence in the calculated lattice parameters for the other 2H-phase TMDs, 
which were obtained using the same methodology. These variations in the crystal structure and atomic mass directly influence the material's phonon modes. To categorize these differences and obtain a distinct Raman fingerprint, we calculated the phonon frequencies and their corresponding Raman activities. Note that the calculated frequencies shown below are based on the LDA method, unless mentioned otherwise. 

\begin{figure*}[ht!]
    \centering
    \includegraphics[width=1.0\textwidth]{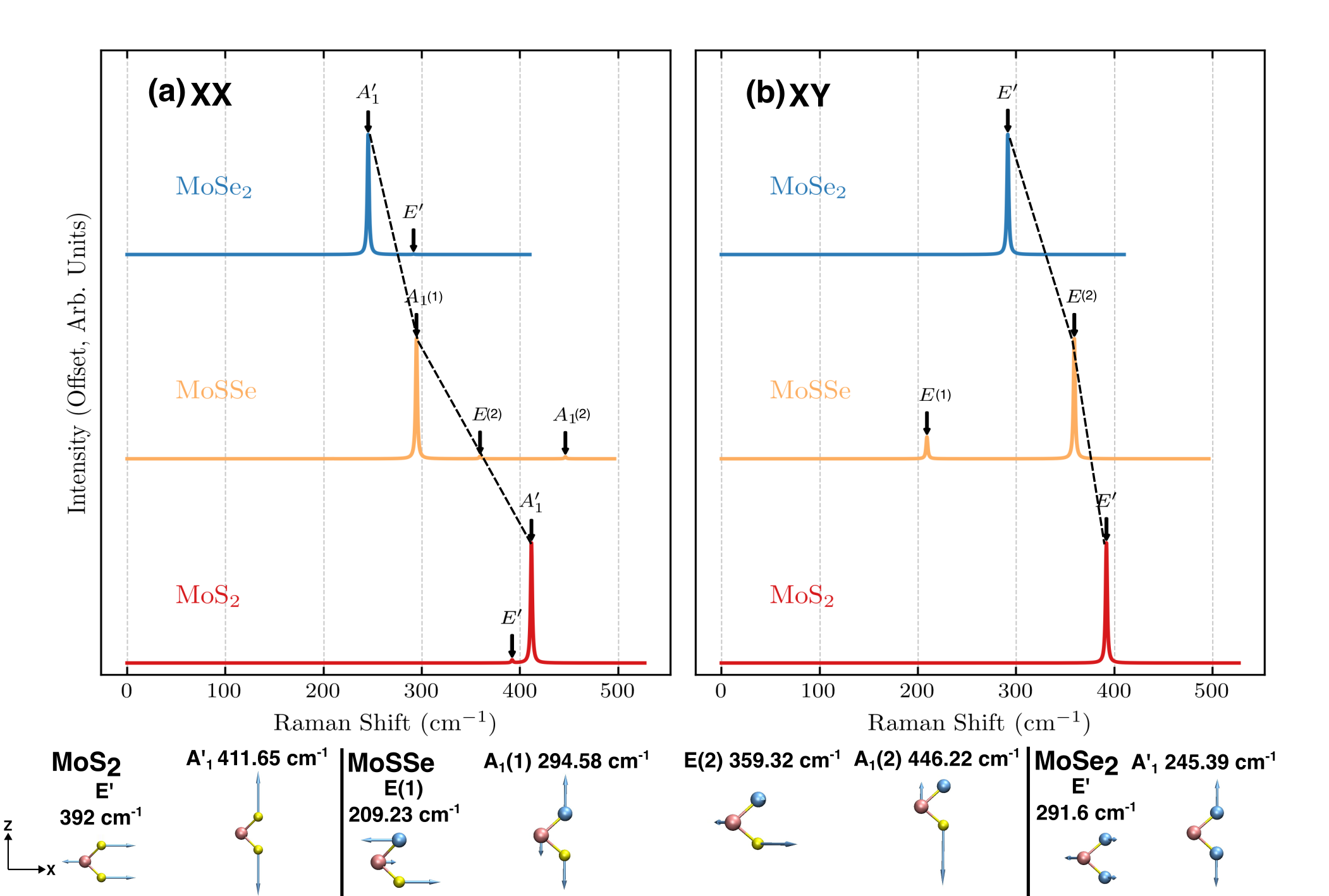}
    \caption{Polarized Raman spectra of 2H-phase monolayer MoS$_2$, MoSSe, and MoSe$_2$ in the (a) XX and (b) XY configurations. 
    The evolution of characteristic modes is shown, with key Raman modes 
    indicated. Corresponding atomic displacements and calculated frequencies are shown on the bottom.}
    \label{fig:mosse-modes}
\end{figure*}

\textbf{The MoSSe family}. Figure \ref{fig:mosse-modes} shows the calculated Raman-active vibrational modes for monolayer MoS$_2$, Janus MoSSe, and MoSe$_2$ as a representative example. We begin with the parent MoS$_2$ monolayer, which belongs to the P6m2 (No. 187) space group with D$_{3h}$ point group symmetry. Group theory predicts the primary first-order Raman-active modes to be the in-plane $E'$ and the out-of-plane $A_1'$ \cite{terronesNewFirstOrder2014}. Our calculations have the $E'$ mode around $392~\mathrm{cm}^{-1}$ and the $A_1'$ mode at approximately $412~\mathrm{cm}^{-1}$, which is consistent with previous theoretical and experimental studies \cite{liangFirstprinciplesRamanSpectra2014, zhouRamanModesMoS22014, molina-sanchezPhononsSinglelayerFewlayer2011}. After the formation of Janus MoSSe monolayer, the out-of-plane mirror symmetry is broken, reducing the crystal symmetry to the P3m1 (No. 156) space group
with a C$_{3v}$ point group. This symmetry reduction alters the Raman selection rules, allowing previously inactive modes to become Raman-active.
In the MoSSe spectrum, the in-plane vibrations are now characterized by two prominent $E$ modes at approximately $209~\mathrm{cm}^{-1}$ and $359~\mathrm{cm}^{-1}$. 
The higher-frequency mode at $359~\mathrm{cm}^{-1}$ is a direct evolution of the parent's Raman-active $E'$ mode. The lower-frequency mode at $209~\mathrm{cm}^{-1}$, 
however, originates from the $E''$ mode of MoS$_2$ that cannot be observed under the back-scattering Raman geometry due to the parent D$_{3h}$ symmetry but becomes observable under the reduced C$_{3v}$ symmetry.
Similarly, the out-of-plane $A_1'$ mode from the parent evolves into the $A_1(1)$ mode at approximately $295~\mathrm{cm}^{-1}$. Additionally, a second 
out-of-plane mode, $A_2''$, which is Raman-inactive in the parent D$_{3h}$ structure, becomes Raman-active under the new C$_{3v}$ symmetry, appearing as 
the $A_1(2)$ peak at $446~\mathrm{cm}^{-1}$ in MoSSe. The emergence of these previously undetected modes is a strong signature of the Janus structure's formation, 
and our calculated results are consistent with previous works \cite{oreshonkovRamanSpectroscopyJanus2022,dengSynthesisJanusMoSSe2025, petricRamanSpectrumJanus2021, linLowEnergyImplantation2020}.

For comparison, the parent MoSe$_2$ monolayer exhibits the identical space group and D$_{3h}$ point group symmetry as MoS$_2$ and therefore features the same Raman-active $E'$ and $A_1'$ modes. We calculate these modes to be at $292~\mathrm{cm}^{-1}$ ($E'$) and $245~\mathrm{cm}^{-1}$ ($A_1'$), which is in agreement 
with experimental reports~\cite{albertl.sinoControllableStructureengineeredJanus2023, soubeletResonanceEffectsRaman2016, smitheNanoscaleHeterogeneitiesMonolayer2018}. This red-shift in frequency 
compared to MoS$_2$ is a well-understood effect attributed to the larger atomic mass of selenium. For completeness, the modes and their frequencies and symmetries are summarized
in Table \ref{tab:phonon_modes_mo_2H}.

\begin{table*}[h]
    \centering
    \caption{Frequencies and symmetries of phonon modes for Mo-based TMDs in the 2H phase.}
    \label{tab:phonon_modes_mo_2H}
    {
    \setlength{\tabcolsep}{3pt}
    \resizebox{0.8\textwidth}{!}{%
    \begin{tabular}{lcccccccc}
        \toprule
        \makecell{\textbf{Phonon} \\ \textbf{Modes}} & \multicolumn{2}{c}{\textbf{Mode \# 1}} & \multicolumn{2}{c}{\textbf{Mode \# 2}} & \multicolumn{2}{c}{\textbf{Mode \# 3}} & \multicolumn{2}{c}{\textbf{Mode \# 4}} \\
        \cmidrule(lr){2-3} \cmidrule(lr){4-5} \cmidrule(lr){6-7} \cmidrule(lr){8-9}
        & freq (cm$^{-1}$) & symm & freq (cm$^{-1}$) & symm & freq (cm$^{-1}$) & symm & freq (cm$^{-1}$) & symm \\
            \midrule
            MoS$_2$ & 289.64 & $E''$ & \textbf{392.00} & \textbf{$E'$} & \textbf{411.65} & \textbf{$A'_{1}$} & 476.96 & $A''_{2}$ \\
            MoSe$_2$ & 170.42 & $E''$ & \textbf{291.60} & \textbf{$E'$} & \textbf{245.39} & \textbf{$A'_{1}$} & 359.97 & $A''_{2}$ \\
            MoTe$_2$ & 119.80 & $E''$ & \textbf{241.46} & \textbf{$E'$} & \textbf{176.49} & \textbf{$A'_{1}$} & 299.77 & $A''_{2}$ \\     
            MoSSe & 209.23 & $E$ & \textbf{359.32} & $E$ & \textbf{294.58} & $A_{1}$ & 446.22 & $A_{1}$ \\
            MoSeTe & 139.75 & $E$ & \textbf{267.08} & \textbf{$E$} & \textbf{203.87} & \textbf{$A_{1}$} & 331.16 & $A_{1}$ \\
            MoSTe & 164.08 & $E$ & \textbf{342.18} & \textbf{$E$} & \textbf{235.54} & \textbf{$A_{1}$} & 421.75 & $A_{1}$ \\
        \bottomrule        
    \end{tabular}
    }
    }
\end{table*}

\begin{figure*}[ht!]
    \centering
    \includegraphics[width=1.0\textwidth]{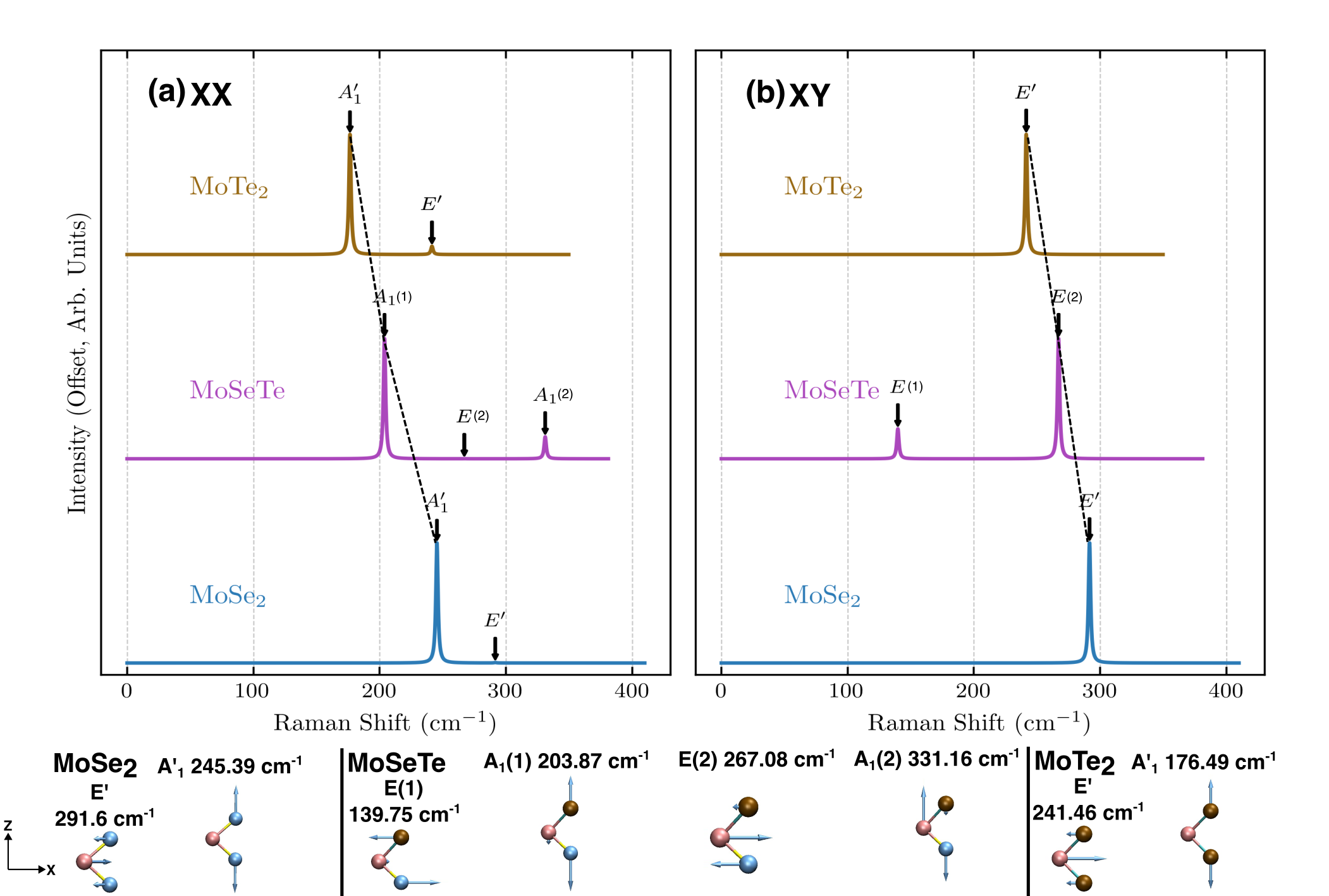}
    \caption{Polarized Raman spectra of 2H-phase monolayer MoSe$_2$, MoSeTe, and MoTe$_2$ in the (a) XX and (b) XY configurations. 
    The evolution of characteristic modes is traced from MoSe$_2$ to MoTe$_2$.}
    \label{fig:mosete-modes}
\end{figure*}

\textbf{The MoSeTe family}. A similar trend in the Raman spectra evolution is seen in the heavier MoSeTe evolution. Figure \ref{fig:mosete-modes} shows the calculated Raman spectra for monolayer MoSe$_2$, Janus MoSeTe, and MoTe$_2$.  
The parent MoTe$_2$ monolayer, which also possesses D$_{3h}$ symmetry and belongs to the P6m2 (No. 187) space group, serves as the starting point. Its characteristic Raman-active $E'$ and $A_1'$ modes 
are found at approximately $241~\mathrm{cm}^{-1}$ and $176~\mathrm{cm}^{-1}$, respectively \cite{grzeszczykRamanScatteringFewlayers2016, guoDoubleResonanceRaman2015}. These frequencies are further red-shifted compared to the sulfide and selenide 
compounds, a direct consequence of the significantly larger atomic mass of tellurium. In the Janus MoSeTe monolayer, the symmetry is again reduced to 
C$_{3v}$, leading to the activation of previously silent modes. The higher-frequency $E$ mode at $267~\mathrm{cm}^{-1}$ and the $A_1$(1) mode at $204~\mathrm{cm}^{-1}$ 
evolve from the parent's active $E'$ and $A_1'$ modes. Crucially, the lower-frequency $E$ mode at $140~\mathrm{cm}^{-1}$ and the $A_1$(2) mode at $331~\mathrm{cm}^{-1}$ 
emerge from the non-observable $E''$ and Raman inactive $A_2''$ parent modes, respectively \cite{yangControllablePhaseModulation2023} (see Table \ref{tab:phonon_modes_mo_2H} for the summary of the modes and their frequencies). The appearance of these additional peaks along with their positions provides a clear indicator for the successful formation of the MoSeTe structure.

\begin{figure*}[ht!]
    \centering
    \includegraphics[width=1.0\textwidth]{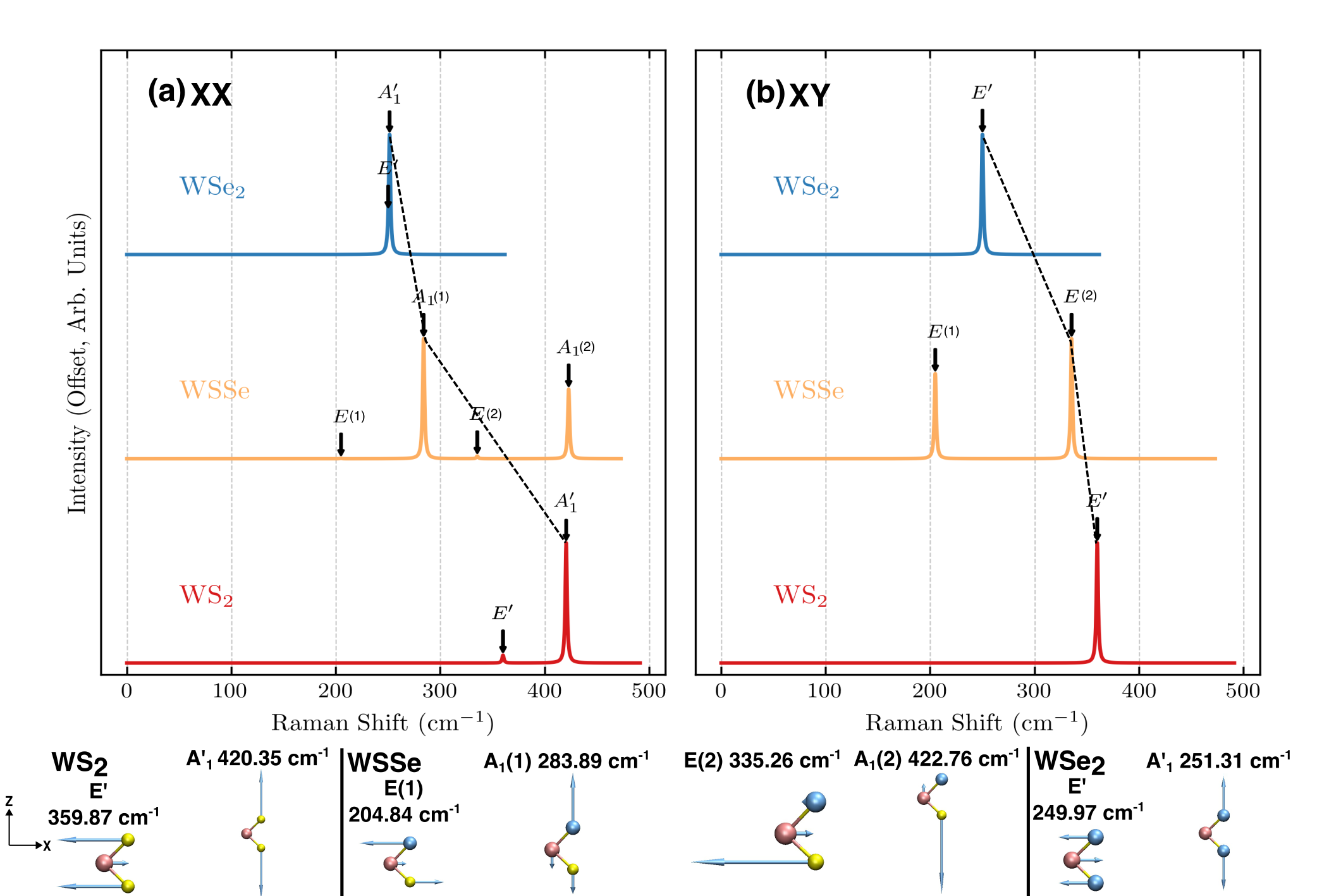}
    \caption{Polarized Raman spectra of 2H-phase monolayer WS$_2$, WSSe, and WSe$_2$ in the (a) XX and (b) XY configurations. The evolution of characteristic vibrational modes is traced from WS$_2$ to WSe$_2$.}
    \label{fig:wsse-modes}
\end{figure*}

\textbf{The MoSTe family}. The last of the Mo-based Janus TMDs analyzed is MoSTe, which also exhibits a similar trend in its Raman spectra. As the properties of the parent MoS$_2$ and 
MoTe$_2$ compounds have already been discussed, we focus here on the signature of the Janus structure shown in Figure S1 in the Supporting Information (SI). Like the 
previously explored Janus structures, MoSTe belongs to the C$_{3v}$ symmetry class and exhibits the characteristic four primary Raman-active peaks.
We identify them at about $164~\mathrm{cm}^{-1}$, $342~\mathrm{cm}^{-1}$, $236~\mathrm{cm}^{-1}$, 
and $422~\mathrm{cm}^{-1}$, corresponding to the $E(1)$, $E(2)$, $A_1(1)$, and $A_1(2)$ modes \cite{yagmurcukardesElectronicVibrationalElastic2019}. The significant mass difference between sulfur and tellurium results in a wide frequency separation between the high-frequency ``S-like'' modes ($E(2)$ and $A_1(2)$) and the 
low-frequency ``Te-like'' modes ($E(1)$ and $A_1(1)$). The same trend is followed for the evolution to the given modes as was for previous Janus structures (see Table \ref{tab:phonon_modes_mo_2H} for the frequencies and symmetries of the modes of MoS$_2$, Janus MoSTe, and MoTe$_2$). This fingerprint, positioned clearly between the spectra of MoS$_2$ and MoTe$_2$, serves as an identifier for the MoSTe monolayer.

\textbf{The WSSe family in the 2H phase}. We now move to the W-based structures in the 2H phase, starting with the Janus WSSe evolution involving WS$_2$ and WSe$_2$. Both WS$_2$ and WSe$_2$ exhibit the same point group symmetry of D$_{3h}$ and P6m2 (No. 187) space group \cite{liangFirstprinciplesRamanSpectra2014}. The calculated Raman spectra
for WSSe are presented in Figure~\ref{fig:wsse-modes}, showing a similar trend to the previous structures. As with its Mo-based counterpart, WSSe adopts 
the C$_{3v}$ point group symmetry, resulting in a characteristic four-peak Raman signature \cite{petricRamanSpectrumJanus2021}. We identify two higher-frequency ``W--S-like'' modes ($E(2)$ and $A_1(2)$) and two 
lower-frequency ``W--Se-like'' modes ($E(1)$ and $A_1(1)$). Specifically, the $A_1$ modes appear at approximately $284~\mathrm{cm}^{-1}$ and $423~\mathrm{cm}^{-1}$, while the $E$ modes are found at $205~\mathrm{cm}^{-1}$ and $335~\mathrm{cm}^{-1}$ \cite{harrisRealTimeDiagnostics2D, linLowEnergyImplantation2020,liuElectronicPropertiesJanus}. These four peaks originate from the evolution of the parent's Raman-active $E'$ and 
$A_1'$ modes, and the previously non-observable $E''$ mode in the back-scattering geometry and the Raman-inactive $A_2''$ mode due to the change of the Raman selection rules, similar to the Mo-based TMDs discussed above. Notably, the frequencies of these modes are slightly different from those in MoSSe, reflecting the influence of the heavier tungsten atom on the lattice dynamics. This unique set of four peaks provides a clear vibrational fingerprint to distinguish WSSe from both its parent compounds (Table \ref{tab:phonon_modes_w_2H}) and its MoSSe analogue. 

\begin{table*}[h!]
    \centering
    \caption{Frequencies and symmetries of phonon modes for W-based TMDs in the 2H phase.}
    \label{tab:phonon_modes_w_2H}
    {
    \setlength{\tabcolsep}{3pt}
    \resizebox{0.8\textwidth}{!}{%
    \begin{tabular}{lcccccccc}
        \toprule
        \makecell{\textbf{Phonon} \\ \textbf{Modes}} & \multicolumn{2}{c}{\textbf{Mode \# 1}} & \multicolumn{2}{c}{\textbf{Mode \# 2}} & \multicolumn{2}{c}{\textbf{Mode \# 3}} & \multicolumn{2}{c}{\textbf{Mode \# 4}} \\
        \cmidrule(lr){2-3} \cmidrule(lr){4-5} \cmidrule(lr){6-7} \cmidrule(lr){8-9}
        & freq (cm$^{-1}$) & symm & freq (cm$^{-1}$) & symm & freq (cm$^{-1}$) & symm & freq (cm$^{-1}$) & symm \\
         \midrule
             WS$_2$ & 298.59 & $E''$ & \textbf{359.87} & \textbf{$E'$} & \textbf{420.35} & \textbf{$A'_{1}$} & 441.06 & $A''_{2}$ \\      
             WSe$_2$ & 175.59 & $E''$ & \textbf{249.97} & \textbf{$E'$} & \textbf{251.31} & \textbf{$A'_{1}$} & 311.69 & $A''_{2}$ \\
             WTe$_2$ & 123.55 & $E''$ & \textbf{200.56} & \textbf{$E'$} & \textbf{181.59} & \textbf{$A'_{1}$} & 250.52 & $A''_{2}$ \\   
             WSSe & 204.84 & $E$ & \textbf{335.26} & \textbf{$E$} & \textbf{283.89} & \textbf{$A_{1}$} & 422.76 & $A_{1}$ \\
             WTeSe & 141.63 & $E$ & \textbf{229.14} & \textbf{$E$} & \textbf{205.47} & \textbf{$A_{1}$} & 286.34 & $A_{1}$ \\
             WSTe & 154.78 & $E$ & \textbf{323.59} & \textbf{$E$} & \textbf{220.77} & \textbf{$A_{1}$} & 400.40 & $A_{1}$ \\           
        \bottomrule
    \end{tabular}
    }
    }
\end{table*}

\textbf{The WTeSe and WSTe families in the 2H phase}. For these W-based TMDs in the 2H phase, similar analysis can be made for the evolution of the frequencies, symmetries, and Raman spectra from the parent to Janus structures (see detailed discussions in Figure S2 and Figure S3 in SI). Table \ref{tab:phonon_modes_w_2H} summaries the frequencies and symmetries of all W-based TMDs in the 2H phase to provide a guiding map for their Raman characterization.

\begin{figure*}[ht!]
    \centering
    \includegraphics[width=1.0\textwidth]{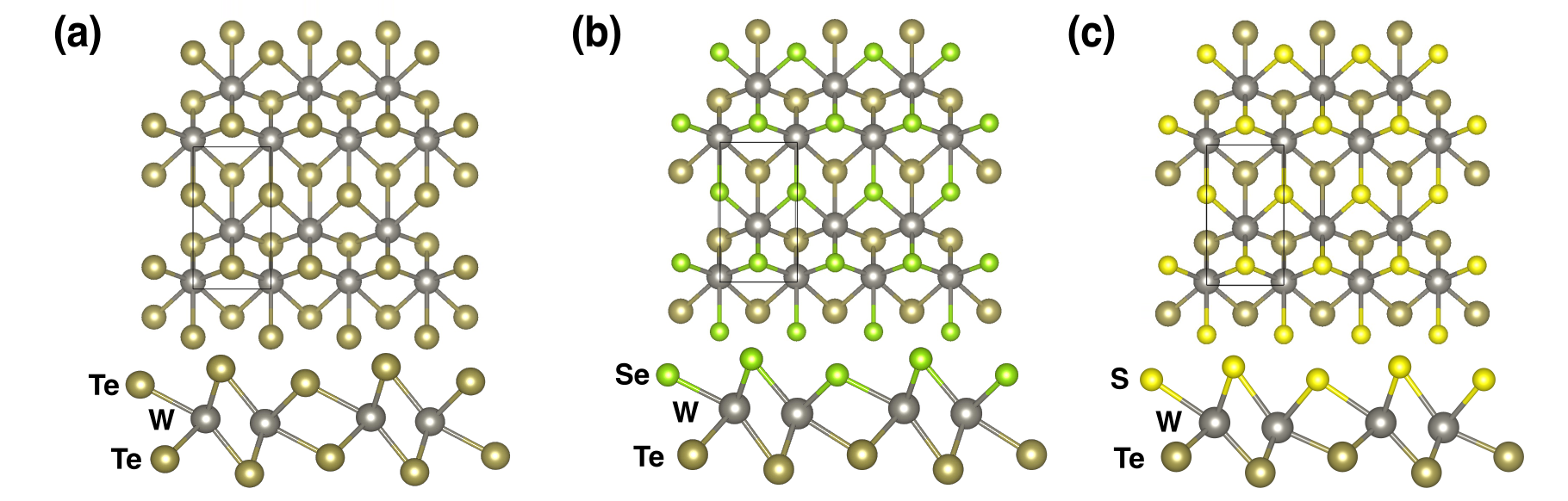}
    \caption{Top and side views of the optimized crystal structures of (a) WTe$_2$, (b) WTeSe, and (c) WSTe monolayers in the Td phase.}
    \label{fig:w-td-structure}
\end{figure*}

\textbf{W-based TMDs in the Td phase}. So far, we have studied the parent and Janus TMDs in the 2H phase since the majority of them are most energetically stable in such a hexagonal phase. However, there are some TMDs that can exist in the orthorhombic Td phase. In particular, for the materials containing the 
W-Te bonds, they can be more energetically stable in the Td phase \cite{caoAnomalousVibrationalModes2017, jiangRamanFingerprintSemimetal2016, ektarawongEffectThermallyExcited2021}. The Td phase is formed through a lattice distortion of the 1T phase, where the transition metal atoms are dislocated to form zigzag chains that doubles the periodicity of the structure (see Figure \ref{fig:w-td-structure} for the atomic structures and Table S1 in SI for the optimized lattice constants). For WTe$_2$, the Td phase gives rise to novel properties that the 2H phase cannot, such as type-II Weyl semimetallic states and  extreme magnetoresistance \cite{Soluyanov2015, aliLargeNonsaturatingMagnetoresistance2014}. As for Janus TMDs, taking WSTe as an example, it can exist as a polymorph in either the 2H phase and Td phase depending on the synthesis method \cite{wang2H1TPhase2020}. Therefore, it is crucial to develop a Raman database for Td-phase TMDs as well to facilitate their identification. 

\begin{figure*}[ht!]
    \centering
    \includegraphics[width=1.0\textwidth]{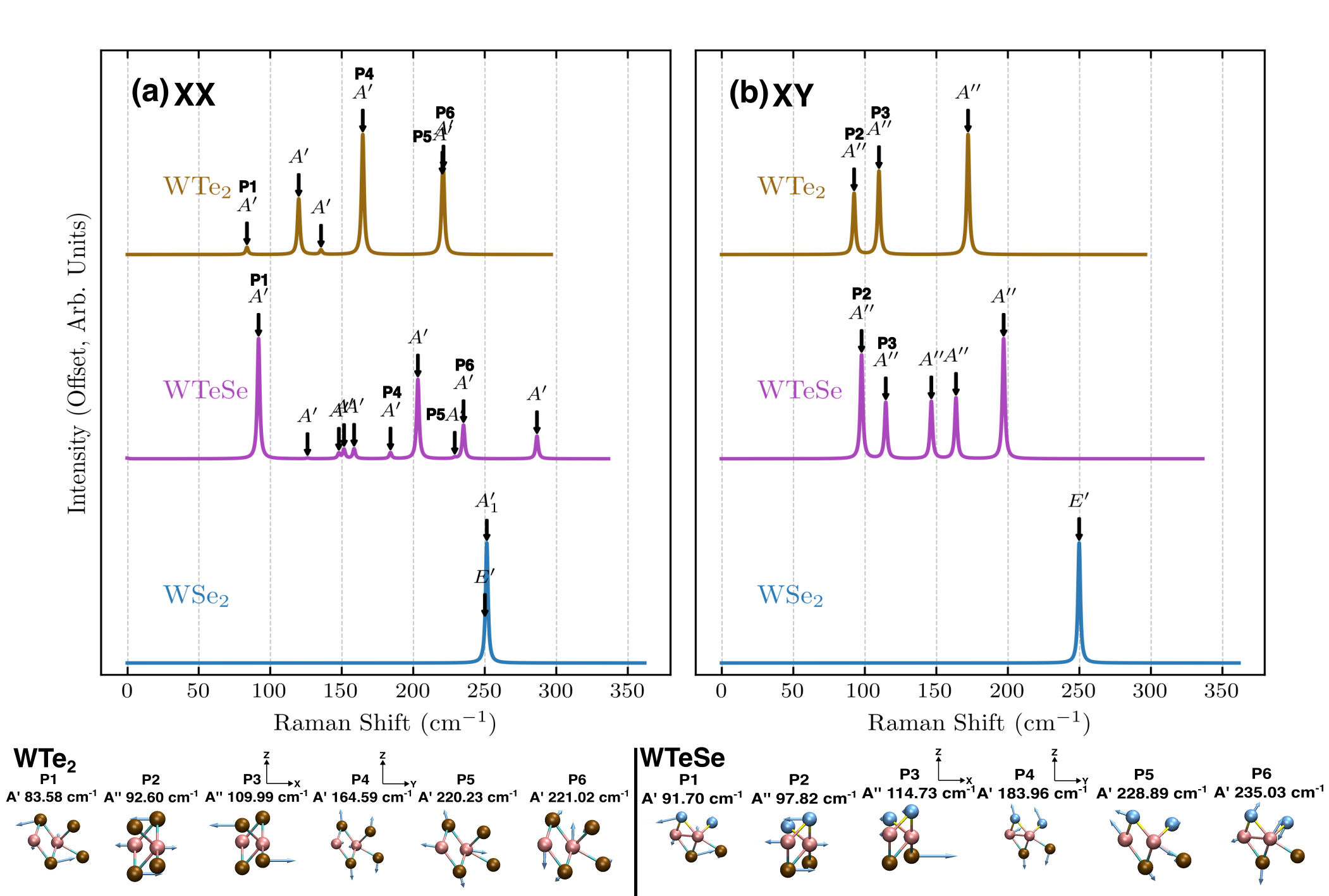}
    \caption{Polarized Raman spectra of monolayer WSe$_2$, WTeSe, and WTe$_2$ in the (a) XX and (b) XY configurations. Corresponding atomic displacement patterns and calculated mode frequencies are shown on the bottom. WTeSe and WTe$_2$ are in the Td phase, while WSe$_2$ is in the 2H phase.}
    \label{fig:wteSe-td-modes}
\end{figure*}

\textbf{WTeSe and WTe$_2$ in the Td phase}. Figure \ref{fig:wteSe-td-modes} shows the calculated Raman spectra for the transition from WTe$_2$ to the Janus WTeSe. Due to the structural distortion and symmetry reduction, Td-TMDs exhibit dramatic differences in the vibrational properties from the high-symmetry 2H-TMDs shown previously. For bulk Td-WTe$_2$, it belongs to the orthorhombic Pmn2$_1$ space group (No. 31); in the monolayer limit, the symmetry becomes Pm (No. 6) with the C$_{s}$ (m) point group. Both the bulk and monolayer structures lack the spatial inversion symmetry. For monolayer WTe$_2$, compared to the 2H phase, the symmetry reduction in the Td phase gives rise to a larger number of Raman-active modes that are dependent on the polarization \cite{songInPlaneAnisotropyWTe22016,zhouPressureinducedTd1T2016, marMetalmetalVsTelluriumtellurium1992,kongRamanScatteringInvestigation2015}. The parallel (XX) spectrum is solely populated by modes with $A'$ symmetry, while the cross (XY) spectrum
is dominated by modes with $A''$ symmetry. We note two near degenerate modes in the XX spectra, modes P5 and P6 in Figure \ref{fig:wteSe-td-modes}, are tricky to tell apart, so the evolution is displayed for both.

\begin{table*}[h!]
    \centering
    \caption{Frequencies and symmetries of phonon mode for TMDs in the Td phase.}
    \label{tab:phonon_modes_Td_wtese}
    {
    \setlength{\tabcolsep}{3pt}
    \resizebox{1.00\textwidth}{!}{%
    \begin{tabular}{lcccccccccccc}
        \toprule
        \makecell{\textbf{Phonon} \\ \textbf{Modes}} & \multicolumn{2}{c}{\textbf{Mode \# 1}} & \multicolumn{2}{c}{\textbf{Mode \# 2}} & \multicolumn{2}{c}{\textbf{Mode \# 3}} & \multicolumn{2}{c}{\textbf{Mode \# 4}} & \multicolumn{2}{c}{\textbf{Mode \# 5}} & \multicolumn{2}{c}{\textbf{Mode \# 6}} \\
        \cmidrule(lr){2-3} \cmidrule(lr){4-5} \cmidrule(lr){6-7} \cmidrule(lr){8-9} \cmidrule(lr){10-11} \cmidrule(lr){12-13}
        & freq (cm$^{-1}$) & symm & freq (cm$^{-1}$) & symm & freq (cm$^{-1}$) & symm & freq (cm$^{-1}$) & symm & freq (cm$^{-1}$) & symm & freq (cm$^{-1}$) & symm \\
        \midrule    
        WTe$_2$  & 83.58 & $A'$ & 92.60 & $A''$ & 109.99 & $A''$ & 164.59  & $A'$ & 220.23 & $A'$ & 221.02 & $A'$ \\           
        WTeSe & 91.70 & $A'$ & 97.82 & $A''$ & 114.73 & $A''$ & 183.96 & $A'$ & 228.89 & $A'$ & 235.03 & $A'$ \\ 
        \midrule 
        \midrule         
        WTe$_2$  & 83.58 & $A'$ & 92.60 & $A''$ & 109.99 & $A''$ & 119.78 & $A'$ & 164.59 & $A'$ & 220.23 & $A'$ \\         
        WSTe & 95.94 & $A'$ & 99.98 & $A''$ & 132.80 & $A''$ & 149.47 & $A'$ & 234.41 & $A'$ & 291.26 & $A'$ \\
        \bottomrule
    \end{tabular}
    }
    }
\end{table*}

The Janus Td-WTeSe also has the Pm (No. 6) space group with C$_s$ point group symmetry \cite{yuMetalSemiconductorPhaseTransition2017}, and it resembles a polar distortion of the Td-phase structure \cite{osti_1652426, PhysRevB.102.060103, 10.1063/1.4959026}. 
Its XX Raman spectrum in Figure \ref{fig:wteSe-td-modes} is characterized by peaks of $A'$ symmetry, while the XY polarization contains the $A''$ modes. Because of the lattice distortion and substantially lower symmetries in these Td structures, the number of Raman peaks is considerably higher than that in 2H structures discussed above, rendering the characterization of the parent and Janus TMDs more challenging. This makes first-principles simulations of these Raman modes and their atomic vibrational patterns more necessary to aid experiments. To trace the evolution from WTe$_2$ to WTeSe in Figure \ref{fig:wteSe-td-modes}, we identified six Raman modes from P1 
to P6 that show clear correspondence between WTe$_2$ to WTeSe in terms of vibrational patterns. For example, the first peak P1, a prominent $A'$ mode in WTe$_2$ at approximately $84~\mathrm{cm}^{-1}$, evolves into the $A'$ mode at around $92~\mathrm{cm}^{-1}$ in WTeSe. 
The mode visualizations confirm that the vibrational pattern is preserved, with the shift of frequency caused by the lighter Se atoms in WTeSe.
Peaks P5 and P6 are of particular interest, as they correspond to the near-degenerate $A'$ modes in WTe$_2$ at around $220~\mathrm{cm}^{-1}$ and $221~\mathrm{cm}^{-1}$ with strong Raman intensities. They are difficult to discern in the spectra of WTe$_2$, but their distinctive vibrational patterns, as shown in the mode visualizations, confirm their separate identities \cite{caoAnomalousVibrationalModes2017}. In WTeSe, they evolve into two distinct $A'$ modes at approximately $229~\mathrm{cm}^{-1}$ and $235~\mathrm{cm}^{-1}$, respectively. The increased separation in peak positions serves as an indicator of the transition to the Janus structure. Another peak of interest is P3, a $A''$ mode in WTe$_2$ at around $110~\mathrm{cm}^{-1}$ with strong signals, which evolves into the $A''$ mode at around $115~\mathrm{cm}^{-1}$ in WTeSe. All six modes and their frequencies are summarized in Table \ref{tab:phonon_modes_Td_wtese}. To the best of our knowledge, this is the first work to detail the vibrational mode and Raman spectrum evolution of Janus WTeSe, providing a valuable reference guiding future experimental studies. Note that the other parent structure, WSe$_2$, is stable in the 2H phase, and thus its Raman spectra are significantly simpler as shown in Figure \ref{fig:wteSe-td-modes}. Because its 2H-phase structure is very different from the Janus Td-WTeSe, we cannot establish effective correlation of Raman modes between the two structures. 

\textbf{Janus WSTe in the Td phase}. As we have already detailed the parent Td-WTe$_2$ structure, we focus on the features of Td-WSTe. It belongs to the Pm (No. 6) space group with C$_s$ (m) point group symmetry \cite{dongLargeInPlaneVertical2017}. As shown in Figure S4 in SI, its complex Raman spectra are similar to those of WTeSe shown in Figure \ref{fig:wteSe-td-modes}, with $A'$ modes dominating the parallel polarization and $A''$ modes in the cross polarization. A direct correspondence
between the vibrational modes of the parent WTe$_2$ and Janus WSTe can be established for six Raman peaks in Figure S4. For instance, the P3 peak in cross polarization, a prominent $A''$ mode in WTe$_2$ at around $110~\mathrm{cm}^{-1}$, evolves into the $A''$ mode at around $133~\mathrm{cm}^{-1}$ in WSTe. Another set of peaks of interest are P5 and P6, which deviate from the aforementioned WTeSe evolution. Here, only one of the near-degenerate $A'$ modes in WTe$_2$ correspond meaningfully to a mode in WSTe: the P6 peak in WTe$_2$ at around $220~\mathrm{cm}^{-1}$ evolves into the P6 peak ($A'$ mode) at approximately $291~\mathrm{cm}^{-1}$ in WSTe. The P5 peak in WTe$_2$, an $A'$ mode at around $165~\mathrm{cm}^{-1}$, evolves into the P5 peak ($A'$ mode) at approximately $234~\mathrm{cm}^{-1}$ in WSTe. This set of peaks should prove useful for identifying the Janus structure. Table \ref{tab:phonon_modes_Td_wtese} displays all six modes and their frequencies showing the evolution from WTe$_2$ to WSTe.


\begin{table*}[h!]
    \centering
    \caption{Predicted phonon frequencies based on the optimal weight parameter for Mo-based TMDs in the 2H phase. Experimental values available in literature are shown for comparison.}
    \label{tab:phonon_modes_mo_2H_weighted}
    {
    \setlength{\tabcolsep}{3pt}
    \resizebox{1.00\textwidth}{!}{%
    \begin{tabular}{lccccccc}
        \toprule
        \makecell{\textbf{Phonon} \\ \textbf{Modes}} & \textbf{Mode \# 1} & \multicolumn{2}{c}{\textbf{Mode \# 2}} & \multicolumn{2}{c}{\textbf{Mode \# 3}} & \multicolumn{2}{c}{\textbf{Mode \# 4}} \\
        \cmidrule(lr){2-2} \cmidrule(lr){3-4} \cmidrule(lr){5-6} \cmidrule(lr){7-8}
        & Calc. (cm$^{-1}$) & Calc. (cm$^{-1}$) & Expt. (cm$^{-1}$) & Calc. (cm$^{-1}$) & Expt. (cm$^{-1}$) & Calc. (cm$^{-1}$) & Expt. (cm$^{-1}$) \\
            \midrule
            MoS$_2$ & 283.81 & \textbf{383.74} & 382--387~\cite{dengSynthesisJanusMoSSe2025, zhouRamanModesMoS22014,  albertl.sinoControllableStructureengineeredJanus2023, liBulkMonolayerMoS22012, golasaResonantRamanScattering2014, linLowEnergyImplantation2020, luJanusMonolayersTransition2017} & \textbf{405.14} & 402--406~\cite{dengSynthesisJanusMoSSe2025, zhouRamanModesMoS22014, albertl.sinoControllableStructureengineeredJanus2023, liBulkMonolayerMoS22012, golasaResonantRamanScattering2014, linLowEnergyImplantation2020, luJanusMonolayersTransition2017} & 468.46 & --- \\
            MoSe$_2$ & 167.28 & \textbf{286.02} & 287--288~\cite{soubeletResonanceEffectsRaman2016, smitheNanoscaleHeterogeneitiesMonolayer2018} & \textbf{241.89} & 238--242~\cite{mahjouri-samaniDigitalTransferGrowth2014, soubeletResonanceEffectsRaman2016, smitheNanoscaleHeterogeneitiesMonolayer2018} & 353.20 & --- \\
            MoTe$_2$ & 117.53 & \textbf{236.63} & 236~\cite{guoDoubleResonanceRaman2015} & \textbf{173.67} & 171~\cite{guoDoubleResonanceRaman2015} & 293.42 & --- \\     
            MoSSe & 205.89 & \textbf{353.03} & 355--357~\cite{dengSynthesisJanusMoSSe2025, albertl.sinoControllableStructureengineeredJanus2023, petricRamanSpectrumJanus2021, linLowEnergyImplantation2020, luJanusMonolayersTransition2017} & \textbf{290.60} & 288--291~\cite{dengSynthesisJanusMoSSe2025, albertl.sinoControllableStructureengineeredJanus2023, petricRamanSpectrumJanus2021, linLowEnergyImplantation2020, luJanusMonolayersTransition2017} & 439.83 & 442--443~\cite{linLowEnergyImplantation2020, petricRamanSpectrumJanus2021} \\
            MoSeTe & 137.21 & \textbf{261.59} & --- & \textbf{200.69} & --- & 324.51 & --- \\
            MoSTe & 161.48 & \textbf{334.95} & --- & \textbf{231.70} & --- & 415.06 & --- \\
        \bottomrule        
    \end{tabular}
    }
    }
\end{table*}

\begin{table*}[h!]
    \centering
    \caption{Predicted phonon frequencies based on the optimal weight parameter for W-based TMDs in the 2H phase. Experimental values available are shown for comparison.}
    \label{tab:phonon_modes_w_2H_weighted}
    {
    \setlength{\tabcolsep}{3pt}
    \resizebox{\textwidth}{!}{%
    \begin{tabular}{lcccccccc}
        \toprule
        \makecell{\textbf{Phonon} \\ \textbf{Modes}} & \multicolumn{2}{c}{\textbf{Mode \# 1}} & \multicolumn{2}{c}{\textbf{Mode \# 2}} & \multicolumn{2}{c}{\textbf{Mode \# 3}} & \multicolumn{2}{c}{\textbf{Mode \# 4}} \\
        \cmidrule(lr){2-3} \cmidrule(lr){4-5} \cmidrule(lr){6-7} \cmidrule(lr){8-9}
        & Calc. (cm$^{-1}$) & Expt. (cm$^{-1}$) & Calc. (cm$^{-1}$) & Expt. (cm$^{-1}$) & Calc. (cm$^{-1}$) & Expt. (cm$^{-1}$) & Calc. (cm$^{-1}$) & Expt. (cm$^{-1}$) \\
            \midrule
            WS$_2$ & 296.63 & --- & \textbf{357.23} & 350--359~\cite{ linLowEnergyImplantation2020, berkdemirIdentificationIndividualFew2013, molasRamanScatteringExcitation2017, harrisRealTimeDiagnostics2D2023} & \textbf{418.37} & 417--419~\cite{linLowEnergyImplantation2020, berkdemirIdentificationIndividualFew2013, molasRamanScatteringExcitation2017} & 438.31 & --- \\
            WSe$_2$ & 174.20 & --- & \textbf{247.84} & 250~\cite{linLowEnergyImplantation2020} & \textbf{249.72} & 247--250~\cite{yangAnharmonicityMonolayerMoS22017, harrisRealTimeDiagnostics2D2023} & 309.07 & --- \\
            WTe$_2$ & 122.53 & --- & \textbf{198.82} & --- & \textbf{180.32} & --- & 248.08 & --- \\   
            WSSe & 202.98 & 207~\cite{petricRamanSpectrumJanus2021} & \textbf{331.79} & 328--335~\cite{petricRamanSpectrumJanus2021, linLowEnergyImplantation2020, harrisRealTimeDiagnostics2D2023} & \textbf{281.50} & 283--286~\cite{petricRamanSpectrumJanus2021, linLowEnergyImplantation2020, harrisRealTimeDiagnostics2D2023} & 419.02 & 420--423~\cite{petricRamanSpectrumJanus2021, linLowEnergyImplantation2020, harrisRealTimeDiagnostics2D2023} \\
            WTeSe & 140.54 & --- & \textbf{227.03} & --- & \textbf{204.06} & --- & 283.74 & --- \\
            WSTe & 153.60 & --- & \textbf{319.96} & --- & \textbf{218.95} & --- & 396.59 & --- \\           
        \bottomrule
    \end{tabular}
    }
    }
\end{table*}

\begin{table*}[h!]
    \centering
    \caption{Predicted phonon frequencies based on the optimal weight parameter for TMDs in the Td phase. Experimental values available are shown for comparison.}
    \label{tab:phonon_modes_Td_weighted}
    {
    \setlength{\tabcolsep}{3pt}
    \resizebox{\textwidth}{!}{%
    \begin{tabular}{lcccccccccccc}
        \toprule
        \makecell{\textbf{Phonon} \\ \textbf{Modes}} & \multicolumn{2}{c}{\textbf{Mode \# 1}} & \multicolumn{2}{c}{\textbf{Mode \# 2}} & \multicolumn{2}{c}{\textbf{Mode \# 3}} & \multicolumn{2}{c}{\textbf{Mode \# 4}} & \multicolumn{2}{c}{\textbf{Mode \# 5}} & \multicolumn{2}{c}{\textbf{Mode \# 6}} \\
        \cmidrule(lr){2-3} \cmidrule(lr){4-5} \cmidrule(lr){6-7} \cmidrule(lr){8-9} \cmidrule(lr){10-11} \cmidrule(lr){12-13}
        & Calc. (cm$^{-1}$) & Expt. (cm$^{-1}$) & Calc. (cm$^{-1}$) & Expt. (cm$^{-1}$) & Calc. (cm$^{-1}$) & Expt. (cm$^{-1}$) & Calc. (cm$^{-1}$) & Expt. (cm$^{-1}$) & Calc. (cm$^{-1}$) & Expt. (cm$^{-1}$) & Calc. (cm$^{-1}$) & Expt. (cm$^{-1}$) \\
        \midrule    
        WTe$_2$  & 81.74 & 79~\cite{kongRamanScatteringInvestigation2015} & 88.85 & 88~\cite{kongRamanScatteringInvestigation2015} & 108.19 & 109--112~\cite{kongRamanScatteringInvestigation2015, jiangRamanFingerprintSemimetal2016} & 160.15 & 160--164~\cite{kongRamanScatteringInvestigation2015, jiangRamanFingerprintSemimetal2016} & 215.33 & 208--212~\cite{kongRamanScatteringInvestigation2015, jiangRamanFingerprintSemimetal2016} & 216.62 & --- \\
        WTeSe & 91.27 & --- & 95.06 & --- & 112.23 & --- & 180.14 & --- & 224.11 & --- & 230.70 & --- \\
        \midrule 
        \midrule         
        WTe$_2$  & 81.74 & 79~\cite{kongRamanScatteringInvestigation2015} & 88.85 & 88~\cite{kongRamanScatteringInvestigation2015} & 108.19 & 109--112~\cite{kongRamanScatteringInvestigation2015, jiangRamanFingerprintSemimetal2016} & 116.93 & 115--120~\cite{kongRamanScatteringInvestigation2015, jiangRamanFingerprintSemimetal2016} & 160.15 & 160--164~\cite{kongRamanScatteringInvestigation2015, jiangRamanFingerprintSemimetal2016} & 215.33 & 208--212~\cite{kongRamanScatteringInvestigation2015, jiangRamanFingerprintSemimetal2016} \\
        WSTe & 95.39 & --- & 97.20 & --- & 129.05 & --- & 145.04 & --- & 230.32 & --- & 284.79 & --- \\
        \bottomrule
    \end{tabular}
    }
    }
\end{table*}

\textbf{Optimizing predictive accuracy.} We note that all the frequencies discussed above are based on the LDA functional. The corresponding results based on the PBE functional are shown in Tables S2-S4 in SI. It is well known that functionals like LDA and PBE exhibit errors, with LDA overbinding and PBE underbinding in general. This suggests that true experimental values generally lie within the range given by the two functionals (i.e., LDA gives the upper limit of the frequencies while PBE yields the lower limit). Since there is no universal functional in DFT calculations that can yield phonon frequencies in perfect agreement with experiments, it is necessary to carry out DFT calculations in both LDA and PBE functionals, thereby presenting a frequency range which experimentalists can use as the reference to compare with experimental values. Obviously, this is still a hurdle for our computational library to enable quick and facile Raman characterization of parent and Janus TMDs. To further improve our predicted results and overcome such a challenge, we compared our LDA and PBE data along with all experimental data we could find in literature, and calculated a weighed average for 2H Mo-based, 2H W-based, and Td W-based structures individually. The model is of the form $\omega_{\text{avg}} = \lambda \cdot \omega_{\text{LDA}} + (1-\lambda) \cdot \omega_{\text{PBE}}$, with $\omega_{\text{LDA}}$ and $\omega_{\text{PBE}}$ as the calculated frequencies reported above and in the SI, respectively, and $\lambda$ is the weighted parameter that determines how much influence each functional has. To determine the optimal weight $\lambda$, we minimize the deviation between our model's predictions ($\omega_{\text{avg}}$) and the experimental frequencies found in literature. For this purpose, we employ the Root-Mean-Square Error (RMSE) which is a common metric for quantifying accuracy among models, defined as $$
\text{RMSE}(\lambda) = \sqrt{\frac{\sum_{i=1}^{N} (\omega_{\text{avg}}(\lambda)_i - \omega_{\text{exp}, i})^2}{N}}.
$$
Here, $\omega_{\text{avg}}(\lambda)_i$ is the frequency predicted by the model for a value $\lambda$ while $\omega_{\text{exp}, i}$ is the experimental frequency for the $i$-th data point, and $N$ is the total number of experimental frequencies we can find in literature. The resulting $\omega_\text{avg}$ yields predictive frequencies better than LDA and PBE on their own, in theory. In Figure S5 in SI, we determined the optimal weight $\lambda$ by finding the minimum RMSE for 2H Mo-based, 2H W-based, and Td W-based structures separately. The subsequent RMSE for each is 2.0627 cm$^{-1}$ (Mo-based), 3.2916 cm$^{-1}$ (W-based), and 3.3849 cm$^{-1}$ (Td-phase), respectively. 

Applying this model to the Mo-based 2H monolayers, we found that an optimal agreement is achieved with an LDA weight of 56\% (Table \ref{tab:phonon_modes_mo_2H_weighted}), suggesting that an even mix of LDA and PBE frequencies provides a good predictive scheme for the set of monolayers. In contrast, for the W-based 2H monolayers, we determined an LDA weight of 78\% as the optimal value as seen in Table \ref{tab:phonon_modes_w_2H_weighted}. This suggests stronger LDA favor for the tungsten-based TMDs. 

Finally, we applied the model to the Td-phase materials but note the lack of experimental data from the Janus structures we present in this work. Therefore, we only have experimental data of WTe$_2$ to benchmark against, which gives us an LDA weight of 57\%, as shown in Table \ref{tab:phonon_modes_Td_weighted}. In short, the optimally weighted frequencies from LDA and PBE results are well consistent with available experimental values for different Raman modes and different TMDs structures from different experimental works, with the deviation generally within $4~\mathrm{cm}^{-1}$ (very often within $2~\mathrm{cm}^{-1}$). This gives us great confidence for the accuracy of our predicted frequencies to guide future Raman characterization of new Janus structures.  

\textbf{Td versus 1T$'$ phase.} Note that for TMDs like WTe$_2$, its distorted 1T structure can assume either Td and 1T$'$ phase. In monolayer form, the structural differences between the two phases are very minor; the key distinction in bulk materials arises from stacking arrangements between layers, which determines if the crystal possesses inversion symmetry \cite{luOriginSuperconductivityWeyl2016, 
maRamanScatteringTransitionmetal2016, feiRobustFerroelectricityMonolayer2016, aliLargeNonsaturatingMagnetoresistance2014, Cheon2021Structural}. The Td phase is non-centrosymmetric, while the 1T$'$ phase is centrosymmetric. For monolayer WTe$_2$, its Td phase belongs to the Pm (No. 6) space group with C$_s$ (m) point group symmetry as discussed above, while the 1T$'$ phase belongs to the P2$_1$/m (No. 11) space group with C$_{2h}$ (2/m) point group symmetry. Although the Raman modes exhibit different irreducible representations between the two phases, the calculated frequencies are almost identical (Figure S6 and Table S5 in SI), underscoring the notion that the two phases are structurally very similar in the monolayer. This explains why Td and 1T$'$ are sometimes used interchangeably in literature. Subtle differences still exist, however, between the two phases for the vibration patterns and Raman intensities of certain modes, as shown in Figure S6. Because of the structural similarity between Td and 1T$'$, it will be difficult for Raman spectroscopy to distinguish them. Nevertheless, it will be an easy task to differentiate between 2H and Td (or 2H and 1T$'$) phases. More importantly, the focus of this work is to integrate the Raman digital twin with Raman spectroscopy to monitor the conversion from the parent to Janus structures, so it is of less importance to definitively know whether the structures are in the Td or 1T$'$ phase.  

\section{Conclusion}

In summary, we have conducted a comprehensive first-principles study of the vibrational properties and Raman spectra of Janus TMD monolayers, specifically focusing on MoSSe, MoSTe, MoSeTe, WSSe, WSTe, and WTeSe. For the 2H phase, by systematically analyzing the evolution of Raman-active phonon modes from their parent TMD compounds (MoS$_2$, MoSe$_2$, MoTe$_2$, WS$_2$, WSe$_2$, and WTe$_2$), we have identified characteristic four-peak signatures unique to each Janus structure. These signatures arise from the combination of high-frequency modes associated with the lighter chalcogen atoms and low-frequency modes linked to the heavier chalcogen atoms. Furthermore, we have explored the Td phase for W-based TMDs, including WTe$_2$ and its Janus derivatives, WTeSe and WSTe, since they can be energetically stable in such a phase. The Td phase is significantly more complicated due to the structural distortion and symmetry reduction, and thus the Raman spectra are complex with considerably higher numbers of polarization-dependent modes. Our atomic-scale simulations reveal the evolution of Raman modes from the parent WTe$_2$ to the Janus WTeSe and WSTe. Moreover, our results show drastic differences in Raman spectra of the same TMDs in 2H and Td phases, enabling the phase identification as well. Our "Raman digital twin" library provides a detailed vibrational and Raman fingerprint for each Janus TMD, facilitating their experimental characterization. The high throughput calculations using different methods and the subsequent data analytics yield phonon frequencies in remarkable agreement with existing experimental data (generally within $4~\mathrm{cm}^{-1}$), demonstrating the accuracy of our database and its predictive power in guiding experiments. Therefore, it allows for a clear distinction between the parent compounds and other byproducts, assists in sample quality assessment, and aids in validating Raman spectroscopic measurements, a crucial step for any potential applications in both laboratory and industry. The computational library is expected to expedite the discovery and development of Janus 2D materials for fundamental research and practical applications.

\begin{acknowledgement}
This work was supported in part by the U.S. Department of Energy, Office of Science,
Office of Workforce Development for Teachers and Scientists (WDTS) under the Science
Undergraduate Laboratory Internships (SULI) program hosted at Oak Ridge National Laboratory 
and administered by the Oak Ridge Institute for Science and Education. Phonon and Raman scattering simulations used resources at the Center for Nanophase Materials Sciences, which is a U.S. Department of Energy Office of Science User Facility at the Oak Ridge National Laboratory. We acknowledge computational resources of the Compute and Data Environment for Science (CADES) at the Oak Ridge National Laboratory, which is supported by the Office of Science of the U.S. Department of Energy under Contract No. DE-AC05-00OR22725. We also used resources of the National Energy Research Scientific Computing Center (NERSC), a DOE
Office of Science User Facility supported by the Office of Science  of the U.S. DOE under
Contract No. DE-AC02-05CH11231 using NERSC award BES-ERCAP0031261. \\

Notice: This manuscript has been authored by UT-Battelle, LLC under Contract No. DE-AC05-00OR22725 with the U.S. Department of Energy.  The United States Government retains and the publisher, by accepting the article for publication, acknowledges that the United States Government retains a non-exclusive, paid-up, irrevocable, world-wide license to publish or reproduce the published form of this manuscript, or allow others to do so, for United States Government purposes.  The Department of Energy will provide public access to these results of federally sponsored research in accordance with the DOE Public Access Plan (http://energy.gov/downloads/doe-public-access-plan).

\end{acknowledgement}

\begin{suppinfo}

The Supporting Information is available free of charge at https:

Raman spectra and vibration patterns of the MoSTe family, the WTeSe family in the 2H phase, the WSTe family in the 2H phase, and the WSTe family in the Td phase; lattice constants of TMDs in the Td phase; calculated frequencies based on the PBE functional; a figure about optimizing weight parameters for data fitting; comparison between Td and 1T$'$ WTe$_2$. 

\end{suppinfo}

\appendix

\bibliography{references}

\end{document}


\clearpage

\section{The MoSTe family}

\begin{figure}[ht!]
    \centering
    \includegraphics[width=1.0\textwidth]{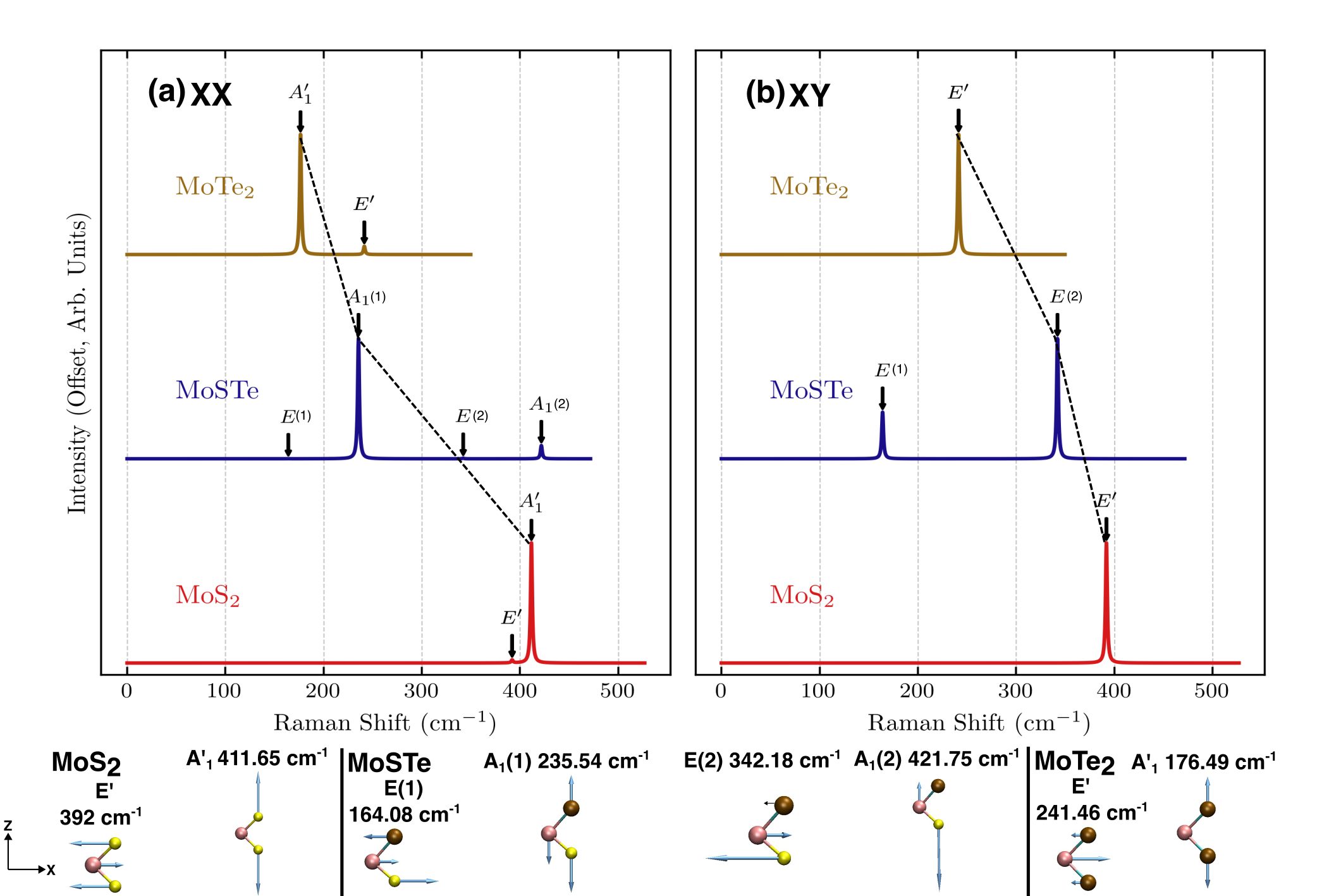}
    \caption{Polarized Raman spectra of 2H-phase monolayer MoS$_2$, MoSTe, and MoTe$_2$ in the (a) XX and (b) XY configurations. The evolution of characteristic vibrational modes is traced from MoS$_2$ to MoTe$_2$, with key Raman-active phonons indicated. Corresponding atomic displacement patterns and calculated mode frequencies are shown on the bottom.}
    \label{fig:moste-modes}
\end{figure}

\clearpage

\section{The WTeSe family in the 2H phase}

\begin{figure}[ht!]
    \centering
    \includegraphics[width=1.0\textwidth]{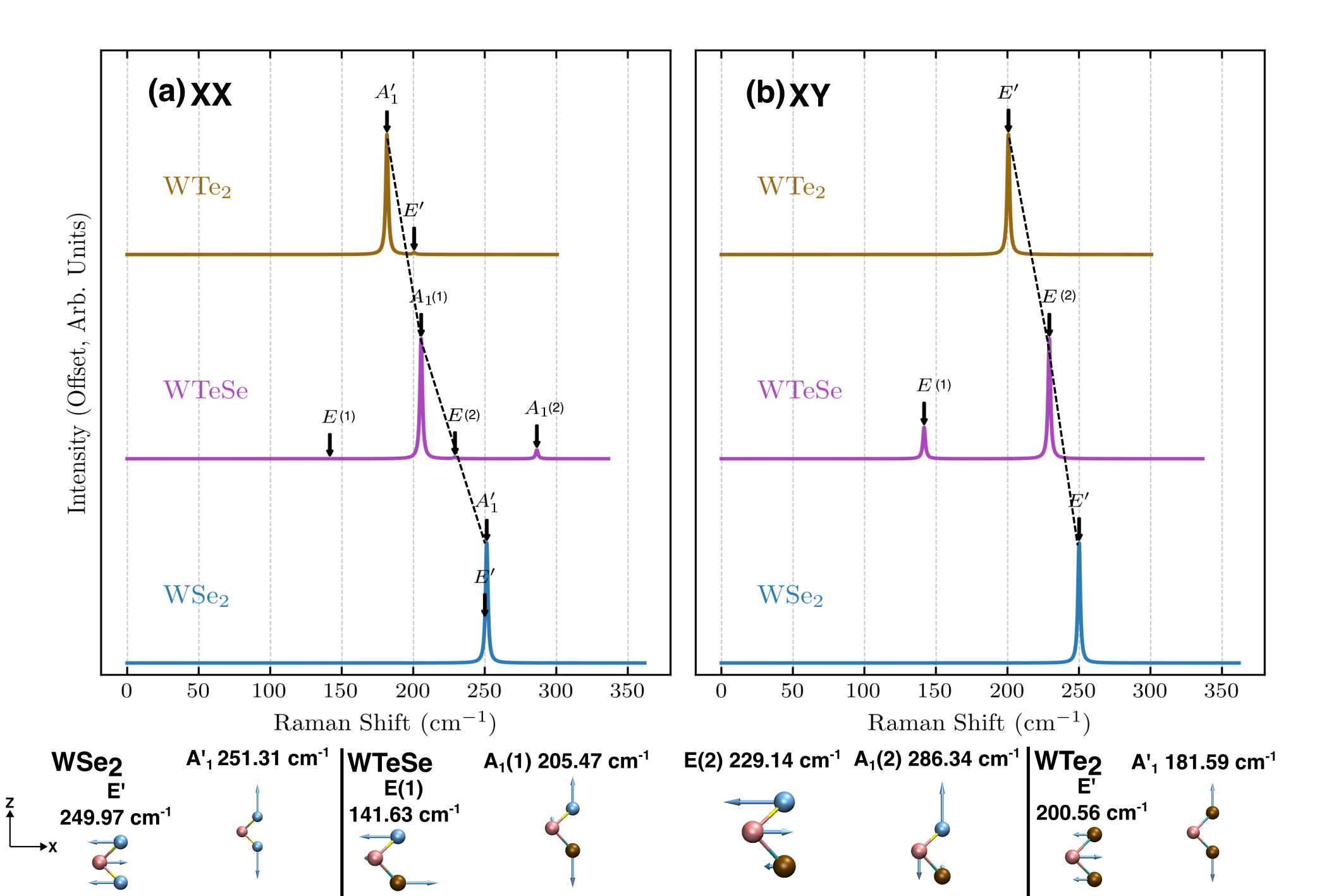}
    \caption{Polarized Raman spectra of monolayer WSe$_2$, WTeSe, and WTe$_2$ in the (a) XX and (b) XY configurations. 
    The evolution of characteristic vibrational modes is traced from WSe$_2$ to WTe$_2$, with key Raman-active phonons 
    indicated. Corresponding atomic displacement patterns and calculated mode frequencies are shown on the bottom. Here all the structures are in the 2H phase.}
    \label{fig:wteSe-2h-modes}
\end{figure}

As with the other Janus structures, WTeSe adopts the C$_{3v}$ point group and a space group of P3m1 (No. 156), which results in a characteristic four-peak signature, as shown in Figure~\ref{fig:wteSe-2h-modes}. We identify two 
higher-frequency ``W--Se-like'' modes and two lower-frequency ``W--Te-like'' modes. Following the established trend, these four peaks arise from the evolution of the parent's Raman-active $E'$ and $A_1'$ modes and the previously Raman non-detectable $E''$ mode in the back-scattering geometry and the Raman-inactive $A_2''$ mode. 

\section{The WSTe family in the 2H phase}

\begin{figure}[ht!]
    \centering
    \includegraphics[width=1.0\textwidth]{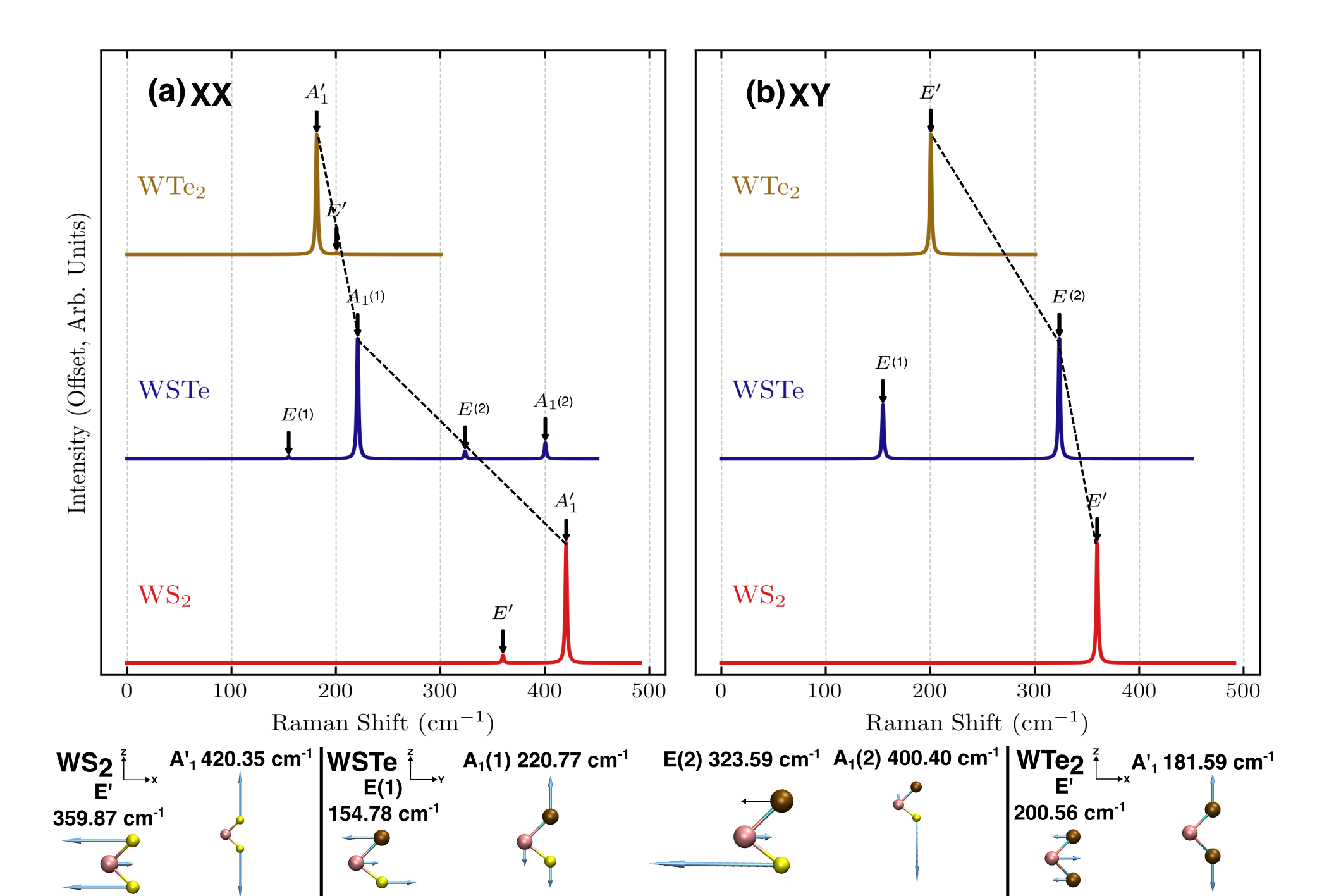}
    \caption{Polarized Raman spectra of monolayer WS$_2$, WSTe, and WTe$_2$ in the (a) XX and (b) XY configurations. 
    The evolution of characteristic vibrational modes is traced from WS$_2$ to WTe$_2$, with key Raman-active phonons 
    indicated. Corresponding atomic displacement patterns and calculated mode frequencies are shown on the bottom. Here all the structures are in the 2H phase.}
    \label{fig:wste-2h-modes}
\end{figure}

The evolution from WS$_2$ to WSTe to WTe$_2$ is shown in Figure~\ref{fig:wste-2h-modes}. WSTe again follows the same trend as the previous structures, exhibiting a characteristic four-peak Raman signature with a space group of P3m1 (No. 156) and a point group symmetry of C$_{3v}$. The $A_1$ modes appear at approximately 
$221~\mathrm{cm}^{-1}$ and $400~\mathrm{cm}^{-1}$, while the $E$ modes are found at $155~\mathrm{cm}^{-1}$ and $324~\mathrm{cm}^{-1}$. Notably, 
the frequencies of these modes are slightly different from those in MoSTe, reflecting the influence of the heavier tungsten atom on the 
lattice dynamics. As with the previous structures, the vibrational mode images allow us to determine which modes correspond to the evolved ones depicted in the spectra figures.
This unique set of four peaks provides a clear vibrational fingerprint to distinguish WSTe from both its parent compounds 
and its MoSTe analogue. 

\section{Lattice constants of TMDs in the Td phase}

\begin{table}[ht!]
    \centering
    \caption{Calculated lattice parameters $a$ and $b$ (\AA) for Td-phase WTe$_2$, WTeSe, and WSTe using LDA and PBE functionals.}
    \begin{tabular}{l|cc|cc}
        \hline
        \multirow{2}{*}{Material} & \multicolumn{2}{c|}{LDA} & \multicolumn{2}{c}{PBE} \\
        & $a$ (\AA) & $b$ (\AA) & $a$ (\AA) & $b$ (\AA) \\
        \hline      
        WTe$_2$ (Td) & 3.4097 & 6.2199 & 3.4907 & 6.3152 \\
        WTeSe (Td)  & 3.3290 & 6.0458 & 3.3965 & 6.1455 \\
        WSTe (Td)   & 3.2760 & 5.9602 & 3.3417 & 6.0622 \\
        \hline
    \end{tabular}
    \label{tab:lattice-params-td}
\end{table}

\clearpage

\section{The WSTe family in the Td phase}

\begin{figure}[ht!]
    \centering
    \includegraphics[width=1.0\textwidth]{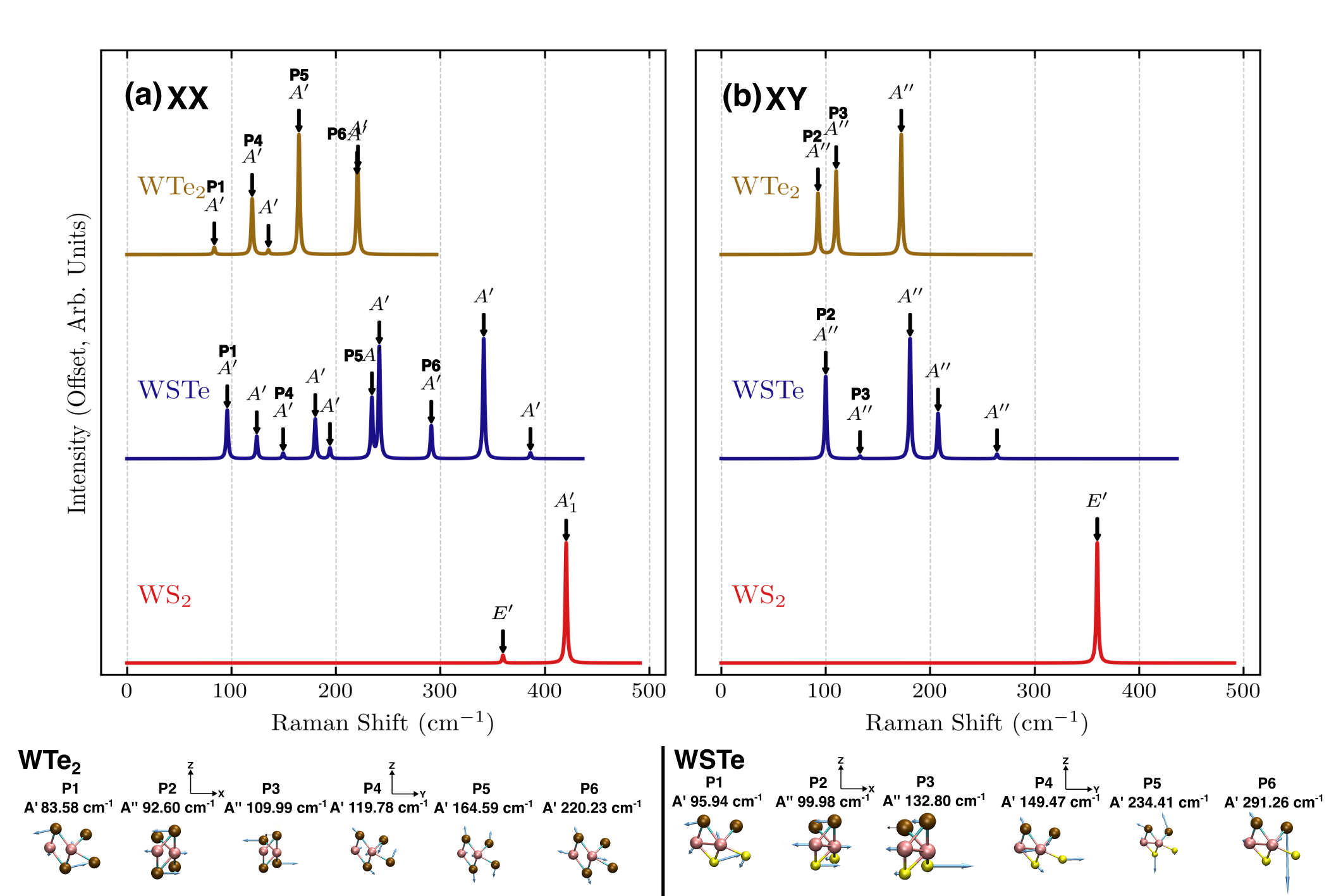}
    \caption{Polarized Raman spectra of monolayer WS$_2$, WSTe, and WTe$_2$ in the (a) XX and (b) XY configurations. 
    Corresponding atomic displacement patterns and calculated mode frequencies are shown on the bottom. WSTe and WTe$_2$ are in the Td phase, while WS$_2$ is in the 2H phase.}
    \label{fig:wste-td-modes}
\end{figure}

\clearpage

\section{Calculated frequencies based on the PBE functional}

\begin{table}[h!]
    \centering
    \caption{Frequencies and Symmetries of Phonon Modes for Mo-based TMDs. They are in the 2H phase. Calculations are based on PBE. The evolution for the Mo-based TMDs using PBE follows the same trend as with LDA reported in the main text. PBE is known to underestimate frequencies, which is shown here when compared to the overestimation of LDA shown in the main text.}
    \label{tab:phonon_modes_mo_2H}
    {
    \setlength{\tabcolsep}{3pt}
    \resizebox{0.95\textwidth}{!}{%
    \begin{tabular}{lcccccccc}
        \toprule
        \makecell{\textbf{Phonon} \\ \textbf{Modes}} & \multicolumn{2}{c}{\textbf{Mode \# 1}} & \multicolumn{2}{c}{\textbf{Mode \# 2}} & \multicolumn{2}{c}{\textbf{Mode \# 3}} & \multicolumn{2}{c}{\textbf{Mode \# 4}} \\
        \cmidrule(lr){2-3} \cmidrule(lr){4-5} \cmidrule(lr){6-7} \cmidrule(lr){8-9}
        & freq (cm$^{-1}$) & symm & freq (cm$^{-1}$) & symm & freq (cm$^{-1}$) & symm & freq (cm$^{-1}$) & symm \\
            \midrule
            MoS$_2$ & 276.38 & $E''$ & \textbf{373.22} & \textbf{$E'$} & \textbf{396.86} & \textbf{$A'_{1}$} & 457.65 & $A''_{2}$ \\
            MoSe$_2$ & 163.29 & $E''$ & \textbf{278.92} & \textbf{$E'$} & \textbf{237.43} & \textbf{$A'_{1}$} & 344.58 & $A''_{2}$ \\
            MoTe$_2$ & 114.63 & $E''$ & \textbf{230.49} & \textbf{$E'$} & \textbf{170.09} & \textbf{$A'_{1}$} & 285.34 & $A''_{2}$ \\     
            MoSSe & 201.63 & $E$ & \textbf{345.02} & $E$ & \textbf{285.54} & $A_{1}$ & 431.70 & $A_{1}$ \\
            MoSeTe & 133.98 & $E$ & \textbf{254.60} & \textbf{$E$} & \textbf{196.64} & \textbf{$A_{1}$} & 316.05 & $A_{1}$ \\
            MoSTe & 158.18 & $E$ & \textbf{325.74} & \textbf{$E$} & \textbf{226.82} & \textbf{$A_{1}$} & 406.54 & $A_{1}$ \\
        \bottomrule        
    \end{tabular}
    }
    }
\end{table}

\begin{table}[h!]
    \centering
    \caption{Frequencies and Symmetries of Phonon Modes for W-based TMDs in the 2H phase, calculated using PBE. The same trend follows for the W-based TMDs using PBE, as with LDA reported in the main text.}
    \label{tab:phonon_modes_w_2H}
    {
    \setlength{\tabcolsep}{3pt}
    \resizebox{0.95\textwidth}{!}{%
    \begin{tabular}{lcccccccc}
        \toprule
        \makecell{\textbf{Phonon} \\ \textbf{Modes}} & \multicolumn{2}{c}{\textbf{Mode \# 1}} & \multicolumn{2}{c}{\textbf{Mode \# 2}} & \multicolumn{2}{c}{\textbf{Mode \# 3}} & \multicolumn{2}{c}{\textbf{Mode \# 4}} \\
        \cmidrule(lr){2-3} \cmidrule(lr){4-5} \cmidrule(lr){6-7} \cmidrule(lr){8-9}
        & freq (cm$^{-1}$) & symm & freq (cm$^{-1}$) & symm & freq (cm$^{-1}$) & symm & freq (cm$^{-1}$) & symm \\
         \midrule
             WS$_2$ & 289.70 & $E''$ & \textbf{347.89} & \textbf{$E'$} & \textbf{411.37} & \textbf{$A'_{1}$} & 428.55 & $A''_{2}$ \\      
             WSe$_2$ & 169.26 & $E''$ & \textbf{240.27} & \textbf{$E'$} & \textbf{244.08} & \textbf{$A'_{1}$} & 299.77 & $A''_{2}$ \\
             WTe$_2$ & 118.90 & $E''$ & \textbf{192.66} & \textbf{$E'$} & \textbf{175.82} & \textbf{$A'_{1}$} & 239.42 & $A''_{2}$ \\   
             WSSe & 196.39 & $E$ & \textbf{319.49} & \textbf{$E$} & \textbf{273.03} & \textbf{$A_{1}$} & 405.77 & $A_{1}$ \\
             WTeSe & 136.67 & $E$ & \textbf{219.57} & \textbf{$E$} & \textbf{199.07} & \textbf{$A_{1}$} & 274.52 & $A_{1}$ \\
             WSTe & 149.40 & $E$ & \textbf{307.07} & \textbf{$E$} & \textbf{212.49} & \textbf{$A_{1}$} & 383.10 & $A_{1}$ \\             
        \bottomrule
    \end{tabular}
    }
    }
\end{table}

\clearpage

\begin{table}[h!]
    \centering
    \caption{Frequencies and Symmetries of Phonon Modes for TMDs in the Td phase, calculated using PBE. For the WTe$_2$ to WTeSe evolution, the same trend holds for PBE like that reported using LDA in the main text. For the WTe$_2$ to WSTe evolution based on PBE, modes 1 and 2 swap for WSTe, with mode 1 leading now.}
    \label{tab:phonon_modes_1T_wtese}
    {
    \setlength{\tabcolsep}{3pt}
    \resizebox{1.00\textwidth}{!}{%
    \begin{tabular}{lcccccccccccc}
        \toprule
        \makecell{\textbf{Phonon} \\ \textbf{Modes}} & \multicolumn{2}{c}{\textbf{Mode \# 1}} & \multicolumn{2}{c}{\textbf{Mode \# 2}} & \multicolumn{2}{c}{\textbf{Mode \# 3}} & \multicolumn{2}{c}{\textbf{Mode \# 4}} & \multicolumn{2}{c}{\textbf{Mode \# 5}} & \multicolumn{2}{c}{\textbf{Mode \# 6}} \\
        \cmidrule(lr){2-3} \cmidrule(lr){4-5} \cmidrule(lr){6-7} \cmidrule(lr){8-9} \cmidrule(lr){10-11} \cmidrule(lr){12-13}
        & freq (cm$^{-1}$) & symm & freq (cm$^{-1}$) & symm & freq (cm$^{-1}$) & symm & freq (cm$^{-1}$) & symm & freq (cm$^{-1}$) & symm & freq (cm$^{-1}$) & symm \\
        \midrule    
        WTe$_2$  & 79.29 & $A'$ & 83.88 & $A''$ & 105.80 & $A''$ & 154.26  & $A'$ & 208.84 & $A'$ & 210.78 & $A'$ \\           
        WTeSe & 90.71 & $A'$ & 91.41 & $A''$ & 108.91 & $A''$ & 175.07 & $A'$ & 217.78 & $A'$ & 224.96 & $A'$ \\ 
        \midrule 
        \midrule         
        WTe$_2$  & 79.29 & $A'$ & 83.88 & $A''$ & 105.80 & $A''$ &  113.16 & $A'$ & 154.26 & $A'$ & 208.84 & $A'$ \\         
        WSTe & 94.67 & $A'$ & 93.52 & $A''$ & 124.07 & $A''$ & 139.17 & $A'$ & 224.90 & $A'$ & 276.22 & $A'$ \\
        \bottomrule
    \end{tabular}
    }
    }
\end{table}

\section{Optimizing weight parameters}

\begin{figure}[ht!]
    \centering
    \includegraphics[width=1.0\textwidth]{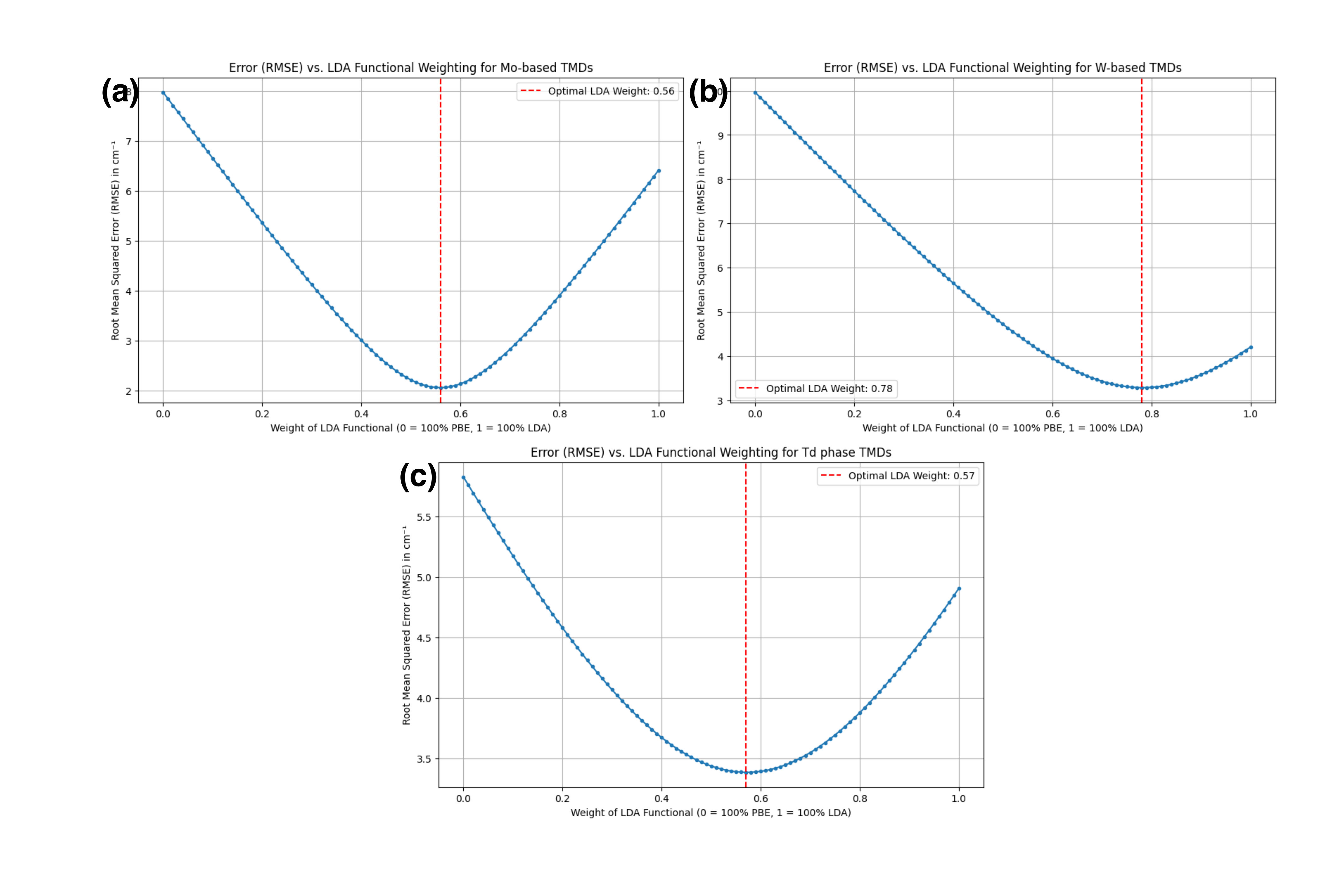}
    \caption{Comparison of experimental data with calculated values using LDA and PBE functionals for (a) 2H phase Mo-based TMDs, (b) 2H phase W-based TMDs, and (c) Td phase TMDs. The subsequent RMSE for each is 2.0627 cm$^{-1}$ (Mo-based), 3.2916 cm$^{-1}$ (W-based), and 3.3849 cm$^{-1}$ (Td-phase) respectively. }
    \label{fig:fittings}
\end{figure}

Displayed are the results from fitting the LDA and PBE calculated frequencies with all experimental data available. The red dashed line represents the minimum of the plotted curve, resulting in the lowest root mean square error (RMSE) at a particular LDA weight.  

\section{Td versus 1T$'$ phase for WTe$_2$}

\begin{figure}[ht!]
    \centering
    \includegraphics[width=1.0\textwidth]{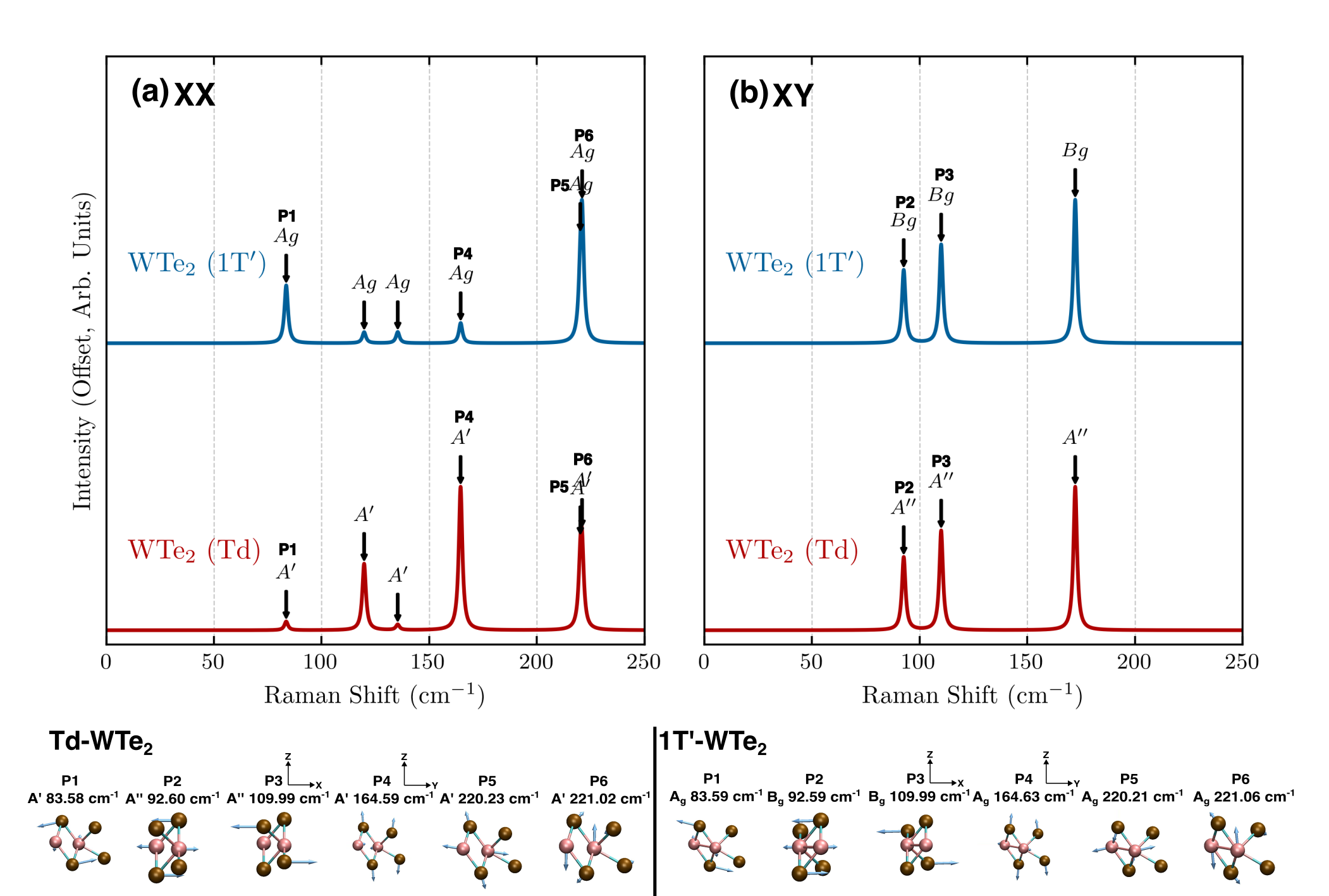}
    \caption{Comparison of Td-phase and 1T$'$-phase WTe$_2$ vibrational modes and Raman intensities in the (a) XX and (b) XY polarized configurations. For some modes, there are differences from vibration patterns and Raman intensities, but their frequencies are almost the same between the two phases.}
    \label{fig:comparison}
\end{figure}

\begin{table}[ht!]
    \centering
    \caption{Comparison of phonon mode frequencies and symmetry labels for WTe$_2$ in the Td and 1T$'$ phases.}
    \label{tab:phonon_modes_Td_expt_vertical}
    \begin{tabular}{lcccc|lccc}
        \toprule
        \textbf{Phase} & \textbf{Mode \#} & \textbf{Calc. (cm$^{-1}$)} & \textbf{Symm.} & & \textbf{Phase} & \textbf{Mode \#} & \textbf{Calc. (cm$^{-1}$)} & \textbf{Symm.} \\
        \midrule
        \multirow{15}{*}{Td} 
        & 1 & 83.58  & $A'$  & & \multirow{15}{*}{1T$'$} 
        & 1 & 83.59  & $A_g$ \\
        & 2 & 92.60  & $A''$ & & & 2 & 92.59  & $B_g$ \\
        & 3 & 109.99 & $A''$ & & & 3 & 109.99 & $B_g$ \\
        & 4 & 114.46 & $A''$ & & & 4 & 114.47 & $A_u$ \\
        & 5 & 119.78 & $A'$  & & & 5 & 119.80 & $A_g$ \\
        & 6 & 131.67 & $A'$  & & & 6 & 131.67 & $B_u$ \\
        & 7 & 132.86 & $A'$  & & & 7 & 132.87 & $B_u$ \\
        & 8 & 135.39 & $A'$  & & & 8 & 135.40 & $A_g$ \\
        & 9 & 158.46 & $A''$ & & & 9 & 158.49 & $A_u$ \\
        & 10 & 164.59 & $A'$  & & & 10 & 164.63 & $A_g$ \\
        & 11 & 172.27 & $A''$ & & & 11 & 172.27 & $B_g$ \\
        & 12 & 180.84 & $A'$  & & & 12 & 180.85 & $B_u$ \\
        & 13 & 220.23 & $A'$  & & & 13 & 220.21 & $A_g$ \\
        & 14 & 221.02 & $A'$  & & & 14 & 221.06 & $A_g$ \\
        & 15 & 246.31 & $A'$  & & & 15 & 246.33 & $B_u$ \\
        \bottomrule
    \end{tabular}
\end{table}